\documentclass[a4paper, 12pt]{article}
\usepackage{a4}
\usepackage{amsfonts}
\usepackage{amssymb}
\usepackage{cite}
\usepackage{epsfig}
\usepackage{psfig}

\author{}

\newcommand{\be}{\begin{equation}}
\newcommand{\ee}{\end{equation}}
\newcommand{\bea}{\begin{eqnarray}}
\newcommand{\eea}{\end{eqnarray}}
\newcommand{\mbb}{\mathbb}

\newcommand{\1}{{\bf 1}}
\newcommand{\3}{{\bf 3}}
\newcommand{\2}{{\bf 2}}
\newcommand{\4}{{\bf 4}}
\newcommand{\5}{{\bf 5}}
\newcommand{\6}{{\bf 6}}

\newcommand{\ov}{\overline}
\def\IR{\relax{\rm I\kern-.18em R}}

\def\IP{\relax{\rm I\kern-.18em P}}
\def\inbar{\vrule height1.5ex width.4pt depth0pt}
\def\IC{\relax\,\hbox{$\inbar\kern-.3em{\rm C}$}}

\def\K3{{\bf K3}}

\def\ov{\overline}

\begin{document}

\title{
\begin{flushright} \vspace{-2cm}
{\small MPP-2005-34 \\
\vspace{-0.35cm}
hep-th/0504232} \end{flushright}
\vspace{4.0cm}
Loop-Corrected Compactifications of the \\
Heterotic String with Line  Bundles  \\ 
\quad
}
\vspace{1.0cm}
\author{\small Ralph~Blumenhagen, Gabriele Honecker and  Timo Weigand}

\date{}

\maketitle

\begin{center}
\emph{Max-Planck-Institut f\"ur Physik, F\"ohringer Ring 6, \\
  80805 M\"unchen, Germany } \\
\vspace{0.2cm}
\tt{blumenha,gabriele,weigand @mppmu.mpg.de}
\vspace{1.0cm}
\end{center}
\vspace{1.0cm}

\begin{abstract}
\noindent  
We consider the $E_8\times E_8$ heterotic string theory compactified on 
Calabi-Yau manifolds
with bundles containing abelian factors in their structure group. 
Generic low energy consequences such as 
the generalised Green-Schwarz mechanism for the multiple anomalous
abelian gauge groups are studied. 
We also compute the holomorphic gauge couplings and induced Fayet-Iliopoulos terms
up to one-loop order, where the latter are interpreted 
as  stringy one-loop corrections
to the Donaldson-Uhlenbeck-Yau condition.
Such models generically have frozen combinations
of K\"ahler and dilaton moduli.
We study concrete bundles with structure group $SU(N)\times U(1)^M$
yielding quasi-realistic gauge groups with
chiral matter given by certain bundle cohomology classes. 
We also provide a number of explicit  tadpole free 
examples  of bundles defined by exact sequences
of sums of line bundles over complete intersection Calabi-Yau  spaces.
This includes one example with precisely the Standard Model
gauge symmetry.

\end{abstract}

\thispagestyle{empty}
\clearpage

\tableofcontents

\section{Introduction}

Over the years string model building techniques have been developed
in various corners of the M-theory moduli space. 
Since one is interested in having gauge symmetry as a general
feature, in particular Type I and heterotic constructions
have been pursued intensively. However, between these
two different constructions there exists a certain asymmetry. 
For Type I like constructions much effort has gone
into the study of so-called intersecting D-brane models,
where Standard Model like features could be engineered
geometrically (see
\cite{AU03,Kiritsis:2003mc,DL04,Blumenhagen:2004vz,Blumenhagen:2005mu} 
for reviews). It is known that in a mirror symmetric picture,
those most simple constructions can be described by 
magnetised D-branes or by turning on abelian gauge fields 
on the world-volume of D9-branes.
On the heterotic side, similar constructions, so-called $(0,2)$ models,
have been investigated, where in most cases people have considered 
non-abelian $SU(N)$ bundles embedded into $E_8\times E_8$
on Calabi-Yau spaces. Various mathematical constructions for
such bundles have been applied, such as the monad construction
occurring naturally in the framework of $(0,2)$ linear sigma
models \cite{Witten:1993yc}, the spectral cover construction 
\cite{Friedman:1997yq} for elliptically
fibered Calabi-Yau manifolds as they appear in studying
F-theory duality and more recently the method
of bundle extensions \cite{Donagi:2000zf,Donagi:2000zs,Donagi:2000fw} 
leading to some MSSM like examples 
\cite{Braun:2005ux,Braun:2005bw} (see also \cite{Curio:1998vu,Andreas:1998ei,
Andreas:1999ty}).
Such $SU(N)$ bundles allow for breaking
the observable gauge symmetry to GUT groups like $E_6$, $SO(10)$
or $SU(5)$. These gauge symmetries can then be broken
further down to the Standard Model group by using Wilson lines. 
Thus for concrete model building, on the Type I side one is
invoking solely $U(1)$ bundles, whereas on the heterotic
side mainly $SU(N)$ bundles have been used.

In this article we  carry out  a more systematic study
of model building prospects using $U(1)$ bundles
in the $E_8\times E_8$ heterotic string 
\footnote{A study   of $U(N)$ bundles in the framework
of the spectral cover construction has appeared  recently
in \cite{Andreas:2004ja}.
Besides that,
the only constructions known to us are 
some scattered results on aspects of  four dimensional models 
\cite{Distler:1987ee,Sharpe:1996xn} and 
a few papers on six-dimensional models 
\cite{Green:1984bx,Strominger:1986uh,Aldazabal:1996du,Berglund:1998va}.}. 
As for magnetised D-brane models, one cannot only break
the gauge symmetry by Wilson lines (trivial line  bundles)
but also by turning on non-trivial (internal) abelian gauge fields. 
The use of these abelian bundles provides much more
model building flexibility and implies some new features
of the resulting models. 

In general, more than one of the $U(1)$ gauge symmetries also
contained in the structure group of the background
gauge bundle is anomalous. We will show in detail that these
anomalies are cancelled by a generalised Green-Schwarz mechanism
invoking not only the universal axio-dilaton multiplet but
also the internal axio-K\"ahler multiplets. Some linear combinations
of the latter become the longitudinal modes of the anomalous
$U(1)$s rendering these gauge fields massive. 
Thus, in contrast  to Wilson line breaking, for the breaking
via non-trivial abelian fluxes the $U(1)$ gauge symmetries
can become massive.
Supersymmetry dictates that there must also arise  Fayet-Iliopolous
terms, which in their supersymmetric minimum give masses to linear combinations of the dilaton
and the K\"ahler moduli. In fact, we will see that the
dilaton part of the FI-terms can be interpreted as a one-loop
correction to the tree-level supersymmetry condition, which is
nothing else than the Donaldson-Uhlenbeck-Yau condition. 
We will provide further evidence for  this picture
from heterotic-Type I duality. Let us emphasize again that in contrast
to earlier claims, $U(1)$ bundles actually freeze combinations
of the axio-dilaton multiplet and the $h_{11}$ axio-K\"ahler multiplets.
As a consequence, such models do not necessarily impose
the condition $h_{11}>1$. 

We also compute the one-loop corrected holomorphic gauge kinetic functions
for both the non-abelian and the abelian gauge factors. 
Including the one-loop threshold corrections, one finds that generically
the gauge couplings for the $U(1)$s are all different.
Therefore, as expected from our original motivation, these heterotic string
models with line bundles show completely analogous features
as their corresponding Type I counterparts. This might help
to resolve some of the puzzles in the literature about
the different structure of heterotic and Type I compactifications \cite{Lalak:1999bk}.

This paper is organised as follows:
In section 2 we  summarize  the general construction of
heterotic string compactifications involving also
$U(N)$  bundles on Calabi-Yau manifolds using (exact) sequences
and discuss 
the generalized Green-Schwarz mechanism cancelling
the various $U(1)$ gauge anomalies. 
Even though there exists an extensive literature on this subject,
for completeness we also discuss the holomorphic gauge kinetic functions. 
As one of the main issues  of this paper, we address the  generation of Fayet-Iliopoulus terms
for the $U(1)$ symmetries,
which together with Type I-heterotic duality provide striking  evidence
for the existence of a stringy one-loop correction to the Donaldson-Uhlenbeck-Yau
stability condition.  
In section 3 we  investigate   bundles
with structure group  $SU(4)\times U(1)$, which give rise
to   
GUT-like
models in four dimensions with
gauge symmetry $SU(5)$. To show that such vacua really exist,
we  provide  a concrete example
of such  bundles both on the Quintic and  on a complete intersection Calabi-Yau (CICY).  
Sections 4 and 5 are devoted to  more general
bundles having two and three  $U(1)$ factors in their structure 
group, for which we 
study the various gauge symmetry enhancements
and the possibility of embedding the MSSM in such a model.
Indeed we exemplify that one can find concrete bundles
which lead  to just the Standard Model gauge symmetry
(in addition to a hidden $E_8$ gauge symmetry). 
Section 6 contains our conclusions. Some technical details are displayed in 
appendix A - C.

\section{Calabi-Yau manifolds with $U(N)$ bundles}

We consider the heterotic string compactified on a Calabi-Yau manifold 
${\cal M}$
endowed  with an additional vector bundle whose structure group is embedded
into the $SO(32)$ or $E_8\times E_8$ ten-dimensional gauge group.
In this paper, for phenomenological reasons, we will be concerned with the
$E_8\times E_8$ heterotic string. In the following the notation
is adjusted  to this case.

\subsection{String model building  constraints} 

Up to now, most papers on heterotic compactifications have considered only bundles
with structure group $SU(N)$.
In this article we investigate  vector bundles of the following
form
\bea
\label{gbundle}
     W=\bigoplus_{i=1}^K  V_{n_i} \oplus \bigoplus_{m=1}^M  L_m,
\eea
where the $V_{n_i}$ are $SU(n_i)$ or $U(n_i)$ bundles
and the $L_m$ denote some complex line bundles with structure
group $U(1)$ each. 
As is well known, to leading order in $\alpha'$ the string equations
of motion respectively the supersymmetry conditions
put several constraints on the vector bundle 
$W$ which can live on the Calabi-Yau. It is one of the main results of this
paper that beyond tree level there arise additional important
constraints at one-loop level which modify the picture in such a 
way as to make it consistent with the dual Type I string constructions.
In the following we move slightly ahead and summarize the main
model building rules arising at string tree and one-loop level.
\begin{itemize}
\item{The vector bundle $W$ has to admit spinors, which means that 
the obstruction given by  the second Stiefel-Whitney class has to vanish, i.e.
\bea
\label{kthe}
       c_1(W)\in H^2({\cal M},2\mbb{Z}).
\eea
}
\item{At string tree level,  the connection of the vector bundle has to satisfy
the well-known zero-slope limit of the Hermitian Yang-Mills equations,
\bea
\label{HYM}
          F_{ab}=F_{\ov a\ov b}=0,\quad     g^{a\ov b}\, F_{a\ov b}=0.
\eea
The first equation implies that 
each term in (\ref{gbundle}) has to be a {\it holomorphic} vector bundle. 
Due to its holomorphicity, this constraint can only arise 
as an F-term in the effective ${\cal N}=1$ supergravity description and therefore 
does not receive any perturbative corrections in $\alpha'$ or the string loop expansion \cite{Dine:1986vd}. 
The second equation in (\ref{HYM}) is actually the special case of the general hermitian Yang-Mills equation
\bea
\label{HYM2}
J \wedge J \wedge F = \mu(W) \, {\rm vol}_{\cal M} \, I ,
\eea
where ${\rm vol}_{\cal M}$ is the volume form on the base manifold of the bundle normalized such that $\int_{\cal M} {\rm vol}_{\cal M} = 1$, and  $I$ refers to the identity matrix acting on the fibre.  Recall that the slope $\mu$ of a coherent sheaf ${\cal V}$ with respect
to a K\"ahler form $J$ on a manifold ${\cal M}$ 
is defined
as  
\bea
      \mu({\cal V})={1\over {\rm rk}({\cal V})} \int_{\cal M} 
          J\wedge J\wedge c_1({\cal V}) .
\eea
According to a theorem by  Uhlenbeck-Yau, (\ref{HYM2}) has a unique solution if and only if the vector bundle $W$ in question is $\mu$-stable, i.e. if  for each coherent subsheaf 
${\cal V}$ of $W$ with 
$0<{\rm rk}({\cal V})<{\rm rk}({W})$
one has
\bea
           \mu ({\cal V})<\mu({W}).
\eea
Consequently, the zero-slope limit of the hermitian Yang-Mills equations {\mbox(\ref{HYM})} relevant at tree level is satisfied precisely by holomorphic $\mu$-stable bundles which meet in addition the integrability condition
\bea
\label{DUY}        \int_{\cal M}  J\wedge J \wedge c_1(V_{n_i}) = 0, \quad\quad
        \int_{\cal M}  J\wedge J \wedge c_1(L_m) = 0,
\eea 
to be satisfied for all $n_i$, $m$. We will refer to the latter constraints in the following as the tree-level Donaldson-Uhlenbeck-Yau
(DUY) equation. 
At string tree level the DUY condition imposes  a set of
constraints on the K\"ahler moduli.
However, we will see in the course of this paper that  the integrability condition (\ref{DUY}) arises from a D-term in the effective
supergravity description, more concretely it is the string tree level
Fayet-Iliopolous term.  Consequently, it is actually a
specific combination of the K\"ahler moduli and the $U(1)$ charged matter fields
which will be frozen by the tree level
requirement of supersymmetry. 
We will however not make the appearance of these matter fields explicit in the 
remainder of this paper.}

\item{ 
As mentioned, the DUY condition plays the role of  
a D-term constraint and as such can, at the perturbative level, in principle be subject to at most one-loop
corrections \cite{Fischler:1981zk}.
We will indeed demonstrate the presence of such 
a one-loop correction to the DUY equation
\bea
\label{DUYloop}
    &&\int_{\cal M} J\wedge J \wedge c_1(L_n)  - \\
    && \frac{1}{2}\, \ell_s^4 \, g_s^2\,
       \int_{\cal M} c_1(L_n) \wedge  \left(
      \sum_{i=1}^K {\rm ch}_2(V_{n_i}) + 
        \sum_{m=1}^M  a_m\, c^2_1(L_m) +{1\over 2}\, c_2(T)\right)=0, \nonumber
\eea
where $g_s=e^{\phi_{10}}$, $\ell_s=2\pi\sqrt{\alpha'}$  and the 
coefficients $a_m\in \mbb{Z}/2$ 
depend on the concrete embedding
of the $U(1)$ structure groups into $E_8\times E_8$ \footnote{Naively one 
might have thought that always $a_m={1\over 2}$, but this is not true. 
Indeed, for the models with $SU(n) \times U(1)^M$ bundles, 
we find that $a_m=6,10,15$ 
if the $U(1)_m$ factor arises via the breaking of the subgroups $E_6 \rightarrow SO(10) \times U(1)$, 
$SO(10) \rightarrow SU(5) \times U(1)$ and $SU(5) \rightarrow SU(3) \times SU(2) \times U(1)$, 
respectively, as we will show in the following sections.}.
Here, the bracket in the second line of~(\ref{DUYloop}) contains only 
sums over those bundles which sit in the same $E_8$ factor as the line bundle $L_n$.
Clearly for $SU(N)$ bundles  the one-loop correction
vanishes just as the tree level constraint and one only gets a non-trivial condition for
$U(N)$ gauge bundles.
The one-loop correction implies that  actually combinations
of K\"ahler moduli and the dilaton  are frozen by the supersymmetry
condition. 
More precisely, if  the $c_1(V_{n_i}), c_1(L_m)\in H^2({\cal M},\mbb{Z})$ 
span a
$D$ dimensional subspace of $H^2({\cal M},\mbb{Z})$, $D$ combinations of the 
K\"ahler moduli and the dilaton become massive.
Observing a modified integrability condition for $U(N)$ bundles, 
there should exist a corresponding
one-loop correction to the hermitian Yang-Mills equation and to the
$\mu$-stability condition.  
}

\item{The Bianchi identity for the three-form 
$H=dB-{\alpha'\over 4}(\omega_{Y}-\omega_L)$,
\bea
\label{TCC}
     0=  dH= {\rm tr}(F_1^2)+{\rm tr}(F_2^2)-{\rm tr}(R^2),
\eea
imposes the so-called tadpole condition for the background bundles.
Here we explicitly distinguish between the observable and hidden gauge sectors and  the trace is over the fundamental representation of $E_8$ for the
gauge bundles and over the fundamental representation of $SO(1,9)$ 
for the curvature two-form. Often one also uses the symbol
${\rm Tr}(F_i)=30\, {\rm tr}(F_i)$ to formally distinguish between traces in the 
adjoint and fundamental representation of $E_8$.
For direct sums of $SU(N)$ bundles the resulting tadpole cancellation
condition takes the familiar form
\bea
\label{taddd}
           \sum_i c_2(V_{n_i}) =c_2(T),
\eea
where $T$ is the tangent bundle of the Calabi-Yau manifold. 
There can be additional non-perturbative contributions to (\ref{taddd})
from wrapped 5-branes, but we are not considering them in this paper
\cite{Lukas:1998hk,Donagi:1998xe,Donagi:1999gc}.
Including abelian bundles, the resulting topological condition can generally  be written as
\bea
\label{TCCgen}
           \sum_{i=1}^K {\rm ch}_2(V_{n_i}) + 
        \sum_{m=1}^M  a_m\, c^2_1(L_m) =-c_2(T).
\eea
Note that the spinor condition (\ref{kthe}) guarantees that
the left hand side takes values in $H^4({\cal M},\mbb{Z})$.
}
\item{In order to finally get a well defined four-dimensional theory, one has
to ensure that the structure group $G$ of the bundle $W$ can be embedded
into $E_8\times E_8$. The observable
gauge group in four dimensions $H$ is the commutant 
of the structure group $G$ of  
$W$ in $E_8\times E_8$. It is clear that the structure group of all line bundles $U(1)^M$
is contained in $H$ ($U(1)$ factors of type (i) according to 
\cite{Dine:1987bq,Distler:1987ee,Lukas:1999nh}), 
but there might be additional  $U(1)$ factors in $H$ not contained
in the structure group ($U(1)$ factors
of type (ii)). }
\end{itemize}

Since $D$  combinations of the dilaton and the K\"ahler moduli become 
massive, supersymmetry implies that the
same must happen to their axionic superpartners as well.
In fact, as we will discuss, $D$ of these axions mix with $D$ of the 
$U(1)$ gauge bosons
making them massive, so that the final gauge symmetry in four dimensions
is reduced by these $D$ abelian $U(1)$ factors.
These massive $U(1)$s nevertheless survive as perturbative
global symmetries providing for instance selection rules
for correlation functions. 
 
The massless spectrum is determined by various cohomology classes
\bea
\label{ghom}
          H^*({\cal M},\bigotimes_{i=1}^K {\textstyle \bigwedge}^{p_i} V_{n_i}\otimes
            \bigotimes_{m=1}^M  L^{q_m}_{m} ),
\eea
where the charges $p_i$ and $q_m$ can  
be derived from the explicit embedding of the structure group into  $E_8\times E_8$.
The net-number of chiral matter multiplets is given by the Euler characteristic of the
respective bundle ${\cal W}$ in (\ref{ghom}), which can be computed using the 
Riemann-Roch-Hirzebruch theorem
\bea 
\label{Euler_character}
         \chi({\cal M},{\cal W})&=&\sum_{i=0}^3  (-1)^i\,  
  {\rm dim}( H^i({\cal M},{\cal W}) ) \nonumber \\
         &=&\int_{\cal M}\left[ {\rm ch}_3({\cal W})+
           {1\over 12}\, c_2(T)\, c_1({\cal W}) \right].
\eea
Of course, allowing this general structure of bundles gives rise
to a plethora of new model building possibilities within the heterotic framework. In the following
we will  discuss some new ways  of how Standard-like
models can arise in this setting. To start with, we focus in this paper 
on three concrete bundle types of the form $V_4\oplus \bigoplus_{m=1}^M  L_m$
and $V_3\oplus \bigoplus_{m=1}^M  L_m$ for up to three $U(1)$ factors.

\subsection{Vector bundles via exact sequences of line bundles}
\label{SubsecVectorbundles}

In order to construct explicit models  one needs a concrete description of vector
bundles on Calabi-Yau spaces.  There are various constructions known in the
literature. For elliptically fibered Calabi-Yau 
spaces one might want to apply the spectral cover construction of Freedman-Morgan-Witten 
\cite{Friedman:1997yq} or the  method of bundle extensions
\cite{Donagi:2000zf,Donagi:2000zs,Donagi:2000fw}.
In this paper we will  however use the definition of vector bundles
as they naturally appear in the $(0,2)$ generalization of the linear sigma
model. Here they are given by (exact) sequences of direct sums of line bundles
on the Calabi-Yau. 

Assume that we have a Calabi-Yau manifold ${\cal M}$ which is given by a complete intersection
in some toric variety and has  $k=h_{11}$ K\"ahler parameters.
A line bundle $L$ on ${\cal M}$ is specified completely by its 
first Chern class which takes
values in $H^2({\cal M},\mbb{Z})$ and can be expanded as
\bea
    c_1(L)=\sum_{i=1}^{h_{11}}  n_i\,  \omega_i,
\eea
where the $\omega_i$ form a basis of $H^2({\cal M},\mbb{Z})$ and 
$n_i \in \mbb{Z}$. 
One also denotes such a line bundle
as ${\cal O}(n_1,\ldots,n_k)$.
Then a vector bundle $V$ of rank $r$ is defined by the cohomology of the monad
\bea
    0\to\ {\cal O}\vert_{\cal M}^{\oplus p}  \stackrel{g}{\longrightarrow} 
         \bigoplus_{a=1}^{r+p+1} {\cal O}(n_{a,1},\ldots, n_{a,k})\vert_{\cal M}
                  \stackrel{f}{\longrightarrow} {\cal O}(m_1,\ldots,m_k)
          \vert_{\cal M}\to 0,
\eea
i.e.  $V=$Kern($f$)/Im($g$) and $p\ge 0$.
Such a sequence can be split into two exact sequences
\bea
   && 0{\longrightarrow} {\cal O}\vert_{\cal M}^{\oplus p} {\longrightarrow}  
      \bigoplus_{a=1}^{r+p+1} {\cal O}(n_{a,1},\ldots, n_{a,k})\vert_{\cal M}
       {\longrightarrow} {\cal E}\vert_{\cal M} {\longrightarrow} 0, \nonumber \\
   &&0{\longrightarrow}\ V\vert_{\cal M} {\longrightarrow}  
         {\cal E}\vert_{\cal M} {\longrightarrow} 
       {\cal O}(m_1,\ldots,m_k)\vert_{\cal M}{\longrightarrow} 0.
\eea
Of course one has to ensure that the maps $g$ and $f$ are such that  the 
sequences really define a bona-fide  vector bundle and not only a non-locally free coherent
sheaf. 

The total Chern class of the vector bundle $V$ is then given by
\bea
\label{totchern}
     c(V)={\prod_{a=1}^{r+p+1} \left (1+\sum_i  n_{a,i}\, \omega_i \right)\over 
           \left (1+\sum_i  m_{i}\, \omega_i\right) },
\eea
which implies in particular that the first Chern class of $V$ is
\bea
     c_1(V)=\sum_{i=1}^{h_{11}} \left(\sum_{a=1}^{r+p+1} n_{a,i} - m_i\right)\, \omega_i .
\eea
Clearly, for $SU(N)$ bundles one has $c_1(V)=0$, whereas for $U(N)$ bundles
at first sight no condition on the first Chern classes arises.
Using the Chern classes one can compute the chiral massless spectrum by means of~(\ref{Euler_character}).
Often also the bundle $\bigwedge^2 V$ appears, whose Chern classes
can be determined  using \cite{Donagi:2004ia} 
\bea
\label{wedgerel}
    {\rm ch}_1({\textstyle \bigwedge^2} V)&=&(r-1)\, {\rm ch}_1(V), \nonumber \\
    {\rm ch}_2({\textstyle \bigwedge^2} V)&=&(r-2)\, {\rm ch}_2(V)+{1\over 2}\, {\rm ch}^2_1(V),   \\
    {\rm ch}_3({\textstyle \bigwedge^2} V)&=&(r-4)\, {\rm ch}_3(V) + {\rm ch}_2(V)\, {\rm ch}_1(V). \nonumber 
\eea
In order to arrive at the complete (non-chiral) massless spectrum 
one really has to compute the cohomology classes $H^i({\cal M},{\cal W})$.
The methods to compute them are known in the literature 
\cite{Distler:1987ee,Distler:1996tj}
and we summarize part of them in Appendix A. 

We pointed out already that there are in principle two distinct ways of embedding abelian groups into $E_8$. 
Either one chooses $V_{n_i}$ to  have structure group $SU(n_i)$. In that case the group theoretic $U(1)$-charges 
of the states upon decomposition of $E_8$ directly give us the powers of the respective line bundles in 
the cohomology class (\ref{ghom}) counting their multiplicities. Clearly, the various line bundles are 
not correlated among each other and in particular $V_{n_i}$ gives no contribution to the $U(1)$-charges. 
We will exemplify this class of constructions by taking the structure group of $V$ to be $SU(4)$ with one and 
two additional line bundles in sections $3$ and $4$, respectively, and by choosing an $SU(3)$ bundle with 
three line bundles in section $5$.

Alternatively, one can embed $U(N)$ bundles into $E_8$ by means of 
a particular construction where one actually starts with a
$U(N)\times U(1)^M$ bundle with $c_1(W)=0$.
 Throughout the remainder of this article, in contrast
to the ansatz~(\ref{gbundle}) for $SU(N) \times U(1)^M$ bundles, we adopt the 
notation 
\bea
\label{gbundle2}
     W=\bigoplus_{i=1}^K  V_{n_i} \oplus \bigoplus_{m=1}^M  L_m^{-1}
\eea
for $U(N)\times U(1)^M$ bundles.
There is a natural way of getting the defining line bundle data of such bundles
from the one of an $SU(N+M)$ bundle. 
Say we have found  already an $SU(N+M)$ bundle $W$ or rather a coherent
sheaf, as for the
coming construction it is not essential that $W$ is locally free. 
Then we can split off $M$ of the line bundles and define
\bea
    L^{-1}_m={\cal O}(n_{m,1},\ldots, n_{m,k})\vert_{\cal M}
\eea
for $m=1,\ldots,M$.
The remaining  sequence now reads 
\bea
    0\to\ {\cal O}\vert_{\cal M}^{\oplus p}  \stackrel{g}{\longrightarrow} 
         \bigoplus_{a=M+1}^{r+p+1} {\cal O}(n_{a,1},\ldots,
         n_{a,k})\vert_{\cal M}
     \stackrel{f}{\longrightarrow} {\cal O}(m_1,\ldots,m_k)\vert_{\cal M}\to 0
\eea
and defines a $U(N)$ bundle $V$ if the maps $f$ and $g$ can be suitably chosen.
Clearly these  bundles satisfy $c_1(V)=\sum_{m=1}^M  c_1(L_m)$ and, 
using (\ref{totchern}), one can show that they  obey  the  constraint
\bea
\label{bundlecon}
    \chi(V)+\sum_{m=1}^M  \chi(L_m^{-1}) = \chi(W).
\eea

In contrast to what we said about $SU(N)$ bundles, now the $U(1)$ charges of 
the states comprise contributions from both the $U(N)$ bundle $V$ and the 
line bundles, which after all are not independent but are chosen just to 
absorb  the diagonal  $U(1)$-charge of $U(N)$ in the splitting  
$ SU(N+M) \rightarrow U(N)\times U(1)^M$. 
Now one has to fix  the embedding of the $U(1)^M$ part
of the structure group into $E_8$ respectively $SU(N+M)$.
For $i=1,\ldots,M$ this can be described  by the charges
\bea
      Q_i=(\underbrace{Q_i(V),\ldots,Q_i(V)}_{N\ {\rm times}},
       Q_i(L_1^{-1}), \ldots, Q_i(L_M^{-1}))
\eea
with
\bea
 N\, Q_i(V)+\sum_{m=1}^M Q_i(L_m^{-1})=0.
\eea
For the detailed computation of the various anomalies associated with the 
$U(1)$-factors, it will turn out to be convenient to introduce the matrix
\bea 
\label{chargematrix}
{\cal Q}_{im} = Q_i(V)+  Q_i(L_m).
\eea
Again, we will make this construction more explicit for a  $U(4) \times U(1)$  and 
$U(4) \times U(1)^2$ bundle as well as a model involving    
 $U(3) \times U(1)^3$ in sections~3-5.

\subsection{$U(1)$ gauge factors and the Green-Schwarz mechanism}
\label{GSM}

As usual in string theory, whereas all irreducible anomalies cancel
directly due to the string consistency constraints \cite{Witten:1984dg} 
such as
tadpole cancellation, the factorisable ones do not.
In four dimensions that means that all non-abelian cubic
gauge anomalies do cancel, whereas the mixed abelian-nonabelian, 
the mixed abelian-gravitational and the cubic abelian ones
do not. As we will discuss in this section, 
they have to be cancelled by a generalised Green-Schwarz
mechanism \footnote{The Green-Schwarz mechanism for several $U(1)$ symmetries in 
$E_8\times E_8$ heterotic compactifications has also been discussed in detail in 
\cite{Lukas:1999nh}, where however the authors have come to  somehow  different
conclusions.}.
 Since each $U(1)$ bundle in the structure group of
the bundle implies a $U(1)$ gauge symmetry in four dimensions,
all these latter three anomalies appear.
We restrict ourselves for brevity to the case that 
$V$ has structure group $SU(N)$; we will
indicate the modifications in the otherwise largely  analogous analysis of $U(N)$ bundles at the end of this section.

Let us write the ten-dimensional gauge fields $F^{10}$ as
$F_i^{10}=F_i+\ov F_i$, where $F_i$ is the external four dimensional
part taking values in $H$ and $\ov F_i$ denotes
the internal six-dimensional part, which
takes values in the structure group $G$ of the bundle. 
Recall that the $U(1)$ factors of type (i) are special as they appear both in
$G$ and $H$.
Computing the field theory mixed $U(1)_m$-$SU(N)^2$ and mixed $U(1)_m$-$G^2_{\mu\nu}$  
anomalies for $m\in\{1,\ldots,M\}$,
one finds that their anomaly six-forms are of the  universal form
\bea
   A_{U(1)_m-SU(N)^2}&\sim &  f_m \wedge {\rm tr} F_1^2\
\left[ \int_{\cal M}  \ov f^m\wedge \left({\rm tr} \ov F_1^2 -
                    {1\over 2} {\rm tr} \ov R^2\right)\right] \label{GSmixedgauge},\\
 A_{U(1)_m-G^2_{\mu\nu}} &\sim&   f_m \wedge {\rm tr} R^2\ 
  \left[ \int_{\cal M}  \ov f^m\wedge  \left({12}\, {\rm tr} \ov F_1^2 - 5\,  
         {\rm tr} \ov R^2 \right)\right] \label{GSmixedgrav}.
\eea
Here we have denoted the four-dimensional $U(1)$ two-form 
field strengths
as $f_m$ and the internal ones  as $\ov f^m$. Moreover, we are here 
considering only line bundles in the first $E_8$ factor, the story for the
second one being completely analogous.  
The $U(1)_m$-$U(1)_n$-$U(1)_p$  
anomalies are slightly more complicated and can be written in the
following general form
\bea
 A_{U(1)_m-U(1)_n-U(1)_p} &\sim& f_m\wedge  f_n\wedge  f_p\,
     \Bigl[ \int_{\cal M}  \ov f^m\wedge \delta_{np} 
            \left( {\rm tr} \ov F_1^2 -
                    {1\over 2} {\rm tr} \ov R^2\right)  + \nonumber\\
         && \phantom{aaaaaaaaaaaaaaa}   c_{mnp} \,
            \ov f^m\wedge \ov f^n\wedge \ov f^p \Bigr] \label{GSpuregauge}.
\eea
Here we have assumed that for at least two $U(1)$s being identical,
the single one is $U(1)_m$. For $m\ne n\ne p$ the first term in
(\ref{GSpuregauge}) is absent. For $n=p$ the relative factor
between the first and the second term in (\ref{GSpuregauge})
can be expressed as
\bea
    c_{mnn}={2 \over 3}\, {\rm tr}_{E_8}(Q_n^2)\, \sigma_{mnn},
\eea
where $Q_n$ is the generator of $U(1)_n$, whose trace is related
to the coefficient $a_n$ in (\ref{TCCgen}) via $4\,a_n={\rm tr}_{E_8}(Q_n^2)$.
The 
$\sigma_{mnn}$ denotes  the symmetry factor of the anomalous diagram,
i.e. $\sigma_{mmm}=1$ and $\sigma_{mnn}=3$ for $m\ne n$.

Let us demonstrate that these anomalies
are cancelled by a generalized Green-Schwarz mechanism
\cite{Green:1984sg} \footnote{Many useful 
formulas on group theoretical identities can be found in \cite{Erler:1993zy}.}.
Throughout this paper we are working in string frame. 
In ten dimensions the gauge anomalies are cancelled by the counter term
\cite{Ibanez:1986xy}
\bea
     S_{GS}= {1\over 48\, (2\pi)^5\, \alpha'}\,  \int  B\wedge X_8,
\eea
where $B$ is the string two-form field and the eight-form
$X_8$ reads, as usual,
\bea   
  X_8={1\over 24} {\rm Tr} F^4 -{1\over 7200} \left( {\rm Tr} F^2\right)^2 
      -{1\over 240} \left( {\rm Tr} F^2\right) \left( {\rm tr} R^2\right)+
       {1\over 8}{\rm tr} R^4 +{1\over 32} \left( {\rm tr} R^2\right)^2 .
\eea
Explicitly taking care of the two $E_8$ factors by writing
$F=F_1+F_2$ one gets
\bea   
  X_8&=&{1\over 4} \left({\rm tr} F_1^2\right)^2+{1\over 4} \left({\rm tr} F_2^2\right)^2
   -{1\over 4} \left({\rm tr} F_1^2\right)\left({\rm tr} F_2^2\right)-
    {1\over 8} \left({\rm tr} F_1^2+{\rm tr} F_2^2\right)\left({\rm tr} R^2\right)+ \nonumber \\
 &&   {1\over 8}{\rm tr} R^4 +{1\over 32} \left( {\rm tr} R^2\right)^2 .
\eea
Using the tadpole cancellation condition~(\ref{TCC}), 
we dimensionally reduce this term to 
\bea
   S_{GS}&=& {1\over 64\, (2\pi)^5\, \alpha'}\, \int B\wedge \left({\rm tr} F_1^2\right)\left({\rm tr} \ov F_1^2
         -{1\over 2} {\rm tr} \ov R^2\right) \label{GS1}\\
    &-&  {1\over 768\, (2\pi)^5\, \alpha'}\, \, \int B\wedge \left({\rm tr} R^2\right)\left( {\rm tr} \ov R^2\right)\label{GS2} \\
   &+& \,{1\over 48\, (2\pi)^5\, \alpha'}\, \int B\wedge \left[{\rm tr}( F_1\ov F_1 )\right]^2  \label{GS3}\\
  &+& {1\over 32\, (2\pi)^5\, \alpha'}\, \, \int B\wedge {\rm tr}( F_1\ov F_1 ) \,
        \left({\rm tr} \ov F_1^2
         -{1\over 2} {\rm tr} \ov R^2\right), \label{GS4}
\eea
where we have shown only the terms for the first $E_8$. 
There are exactly the same terms (\ref{GS1}),(\ref{GS3}),(\ref{GS4}) for the second $E_8$ by replacing
$F_1\to F_2$ and $\ov F_1\to \ov F_2$ as well as a mixed term involving $\mbox{tr}(F_1 \ov{F}_1) \, \mbox{tr}(F_2 \ov{F}_2)$. Note, however, that our models in this article only involve line bundles in one of the two $E_8$-factors. 
For concreteness let us discuss the $U(1)$-$SU(N)^2$ and
$U(1)$-$G^2_{\mu\nu}$ anomalies in detail.
It is convenient to make use of  a basis of two-forms $\omega_k$,
 $k=1,\ldots, h_{11}$ as in section~\ref{SubsecVectorbundles}
and their Hodge dual
four-forms  $\widehat{\omega}^k$, i.e. they satisfy
\bea
      \int_{\cal M} \omega_k\wedge \widehat{\omega}_{k'}=\delta_{kk'}.
\eea
In terms of the string length ${\ell}_s=2\pi\sqrt{\alpha'}$ we now expand 
\bea
    B^{(2)}&=& b^{(2)}_0+\ell_s^2\, 
          \sum_{k=1}^{h_{11}}   b^{(0)}_k\, \omega_k ,\quad
    {\rm tr} \ov F_1^2
= (2\pi)^2\, \sum_{k=1}^{h_{11}} 
({\rm tr} \ov F_1^2)_k
\, \widehat{\omega}_{k}, \nonumber \\
   \ov f^m&=&2\pi\, \sum_{k=1}^{h_{11}}   \ov f^m_k\, \omega_k ,\quad\phantom{aaaa}
     {\rm tr} \ov R^2=(2\pi)^2\, \sum_{k=1}^{h_{11}} 
      \bigl( {\rm tr} \ov R^2\bigr)_k\, \widehat{\omega}_{k}, 
\eea
where for dimensional reasons we have introduced appropriate
powers of $\alpha'$. Note that $ \ov f^m_k\in \mbb{Z}$. 
Inserting these expansions into
the GS-terms (\ref{GS1}) and (\ref{GS2})  gives rise to GS-terms in
four dimensions
\bea
 S_{GS}&=& {1\over 64\, (2\pi)}
                 \, \int_{\IR_{1,3}}  \sum_{k=1}^{h_{11}}   
       \left( b^{(0)}_k \, {\rm tr} F_1^2 \right) 
         \,\, \bigl({\rm tr} \ov F_1^2
         -{1\over 2} {\rm tr} \ov R^2\bigr)_k \label{fGSa}\\
     && -{1\over 768\, (2\pi)} \,
       \int_{\IR_{1,3}}  \sum_{k=1}^{h_{11}}   
       \left( b^{(0)}_k \, {\rm tr} R^2 \right) 
         \,\, \bigl({\rm tr} \ov R^2\bigr)_k.\label{fGSb}
\eea
As indicated, from the four-dimensional point of view the 
axions $b^{(0)}_k$ are zero-forms.
However, in addition the term (\ref{GS4}) gives rise to a mass term
for the four-dimensional two-form field $b^{(2)}_0$
\bea
\label{massa}
 S^0_{mass}= {1\over 32\, (2\pi)^5 \alpha'} \, 
            \int_{\IR_{1,3}}  \sum_{m=1}^M   
       \left( b^{(2)}_0 \wedge f_m \right)\, {\rm tr}_{E_8}(Q^2_m)\, 
     \int_{\cal M} \ov f^m\wedge
        \bigl({\rm tr} \ov F_1^2
         -{1\over 2} {\rm tr} \ov R^2\bigr),
\eea
 where we have assumed that ${\rm tr}_{E_8}(Q_m\, Q_n)=0$ for $m\ne n$, which
is indeed satisfied for all $U(1)$ symmetries discussed in this article. 
This mass term for the universal axion is only present
for $U(1)$ symmetries of type (i), reflecting the fact that
for the $E_8\times E_8$ heterotic string $U(1)$ factors
of type (ii) are always non-anomalous. 

To cancel the anomalies one needs a  GS-term for the
external axion $b^{(2)}_0$ and  mass terms for the internal ones $b^{(2)}_k$.
All these terms arise from the
following kinetic term in the 10D effective action
\bea
\label{kinH}
         S_{kin}=-{1\over 4\kappa_{10}^2}\, \int e^{-2\phi_{10}}\,
       H\wedge \star_{10}\, H,
\eea
where $\kappa^2_{10}={1\over 2}(2\pi)^7\, (\alpha')^4$ and
the heterotic 3-form field  
strength reads $H=dB^{(2)}-{\alpha'\over 4}(\omega_{Y}-\omega_{L})$ 
involving the 
gauge and gravitational Chern-Simons terms.
Let us denote the dual 6-form of $B^{(2)}$ as $B^{(6)}$, i.e.
$\star_{10}\, dB^{(2)}=e^{2\phi_{10}}\,dB^{(6)}$. 
Then the essential term contained
in (\ref{kinH}) is
\bea
\label{kinem}
  S_{\rm kin}&=& {\alpha'\over 8\kappa_{10}^2}\,
       \int ({\rm tr}  F_1^2
         - {\rm tr}  R^2 )\wedge B^{(6)} .
\eea
We introduce the dimensionally reduced zero- and two-forms 
\bea
\label{reduce}
      B^{(6)}=\ell_s^6\, b^{(0)}_0\,  {\rm vol}_6 +
      \ell_s^4\,
           \sum_{k=1}^{h_{11}}  b^{(2)}_k\, \widehat\omega_k,
\eea
where ${\rm vol}_6$ is the normalized volume form on ${\cal M}$, i.e
$\int_{\cal M} {\rm vol}_6=1$.  The two-forms $b^{(2)}_k$ in (\ref{reduce})
can be proven to satisfy $\star_{4}\, db_k^{(2)}=e^{2\phi_{10}}\,db_k^{(0)}$
for all $k\in\{1,\dots,h_{11}\}$.
Then the first term in (\ref{reduce}) and equation 
(\ref{kinem})  give first rise to
a four-dimensional GS-term
\bea
\label{fGSc}
  S^0_{GS}= {1\over 8\pi} \int_{\IR_{1,3}} b^{(0)}_0\wedge ({\rm tr}  F_1^2
         - {\rm tr}  R^2 ) ,
\eea
where $\star_{4}\, db_0^{(2)}=e^{2\phi_{10}}\,db_0^{(0)}$.
In addition, reducing such that $F$ takes values
in the various $U(1)$s with one factor external   and 
the other one internal one finds mass terms for the 
various internal axions.  
Performing the dimensional reduction we eventually arrive at four-dimensional
couplings of the form
\bea
\label{massb}
        S_{\rm mass}={1\over 2 \ell^2_s }\,
         \int_{\IR_{1,3}} \sum_{m=1}^M \sum_{k=1}^{h_{11}}  
       \left( f_m\wedge b^{(2)}_k\right)\, {\rm tr}_{E_8}(Q^2_m)\,
          \ov f^m_{k}.
\eea

\begin{figure}
\begin{center}
\epsfxsize=3.5in 
\epsfbox{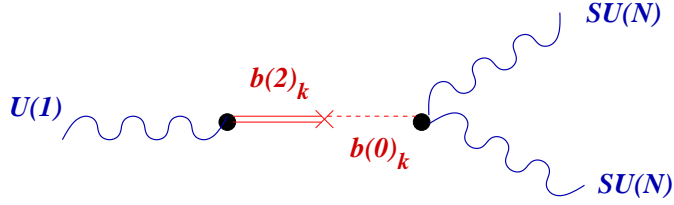}
\caption{Green-Schwarz counterterm for the mixed gauge anomaly.
\label{figGS}}
\end{center}
\end{figure}

The two  GS-couplings (\ref{fGSa},\ref{fGSc})
and two mass terms (\ref{massa},\ref{massb}) have precisely
the right form to generate tree-level graphs of the form
shown in Figure~\ref{figGS}, which provide couplings of the same
type as the ones appearing in the mixed gauge anomalies.
For the mixed-abelian non-abelian GS contribution we get
\bea
      A^{GS}_{U(1)_m-SU(N)^2}\sim   {{\rm tr}_{E_8}(Q^2_m)\over 64 (2\pi)^6\, \alpha'}\, 
               f_m\wedge {\rm tr} F_1^2\ \left[
      \int_{\cal M}  \ov f^m\wedge \left({\rm tr} \ov F_1^2
         -{1\over 2} {\rm tr} \ov R^2\right) \right].
\eea
For the mixed-abelian-gravitational anomaly the contributions
from internal axions and the four-dimensional one
are different but they do add precisely up to
\bea
      A^{GS}_{U(1)_m-G^2_{\mu\nu}}&\sim &  -{{\rm tr}_{E_8}(Q^2_m)
         \over 128 (2\pi)^6\, \alpha'}\,
                f_m\wedge {\rm tr} R^2\ \biggl[
      \int_{\cal M}  \ov f^m\wedge \left({\rm tr} \ov F_1^2
         -{1\over 2} {\rm tr} \ov R^2\right) \nonumber \\
      &&\phantom{aaaaaaaaaaa}    +{1\over 12}  \int_{\cal M}  \ov f^m\wedge 
            \left({\rm tr} \ov R^2\right)  \biggr]\\
     &=& -{{\rm tr}_{E_8}(Q^2_m)\over 128 (2\pi)^6\, \alpha'}\,
                f_m\wedge {\rm tr} R^2\ \left[
      \int_{\cal M}  \ov f^m\wedge \left({\rm tr} \ov F_1^2
         -{5\over 12} {\rm tr} \ov R^2\right)  \right].\nonumber
\eea

Along the same lines,
one can also show that the mixed $U(1)^3$  anomalies cancel, where
here   
also the Green-Schwarz couplings (\ref{GS3}) contribute.

Finally, let us discuss the mass terms (\ref{massa},\ref{massb}) 
in more detail.
Independently of the number of anomalous $U(1)$ gauge factors,
the number of massive $U(1)$ gauge fields is given by the
rank of the matrix 
\bea
    {\rm M}_{mk}=\cases{ {{\rm tr}_{E_8}(Q^2_m)\over 2 \ell_s^2}\, \ov f^m_{k} & for $k\in\{1,\ldots,h_{11}\}$ \cr
                      {{\rm tr}_{E_8}(Q^2_m)\over 32 \ell_s^2 }
              \int_{\cal M}  {\ov f^m\over (2\pi) }\wedge
\, {\left({\rm tr} 
                       \ov F_1^2 -{1\over 2} {\rm tr} \ov R^2\right)\over (2\pi)^2} &
                   for $k=0$. \cr}  
\eea
Let us point out that all mass terms are of the same order in both string
and sigma model perturbation theory. 
The  number of massive $U(1)$s will always be at least as big as
the number of anomalous $U(1)$s, but as in the 
Type I case \cite{ibanez}, 
in principle their numbers  do not necessarily have to agree.
Though all entries in the mass matrix are of order $M_s^2$, 
the mass eigenstates  of the gauge bosons can have
masses significantly lower than the string scale.
  
Analogously, the number of massive  axions $b^{(0)}_i, i=0,\ldots, h_{11}$,
is given by the rank of ${\rm M}$.
Since these axions are just the fields which complexify
the K\"ahler moduli and the dilaton, supersymmetry dictates that the same
number of the latter  should also get a mass.
However, the only source for mass terms so far is the  tree-level DUY equation. 
Expand 
\bea
\label{KaehlerformExpansion}
J=\ell_s^2\,\sum_i \alpha_i\, \omega_i
\quad \mbox{and} \quad
J\wedge J=\ell_s^4\, \sum_i \sigma_i\, \widehat \omega_i, 
\eea
where
$\sigma_i=\sum_{j,k=1}^{h_{11}} d_{ijk}\alpha_j\alpha_k$ and 
$d_{ijk}=\int_{\cal M} \omega_i \wedge \omega_j \wedge \omega_k$  denotes the
triple intersection number.
Then the tree-level DUY condition can be written as 
\bea
    \int_{\cal M} J\wedge J\wedge c_1(L_m)=\ell_s^4 \,
        \sum_{k=1}^{h_{11}}  \sigma_k\, \ov f^m_k=0,
\eea
which tells us that certain linear combinations of the
K\"ahler moduli  become massive. 
However, the dilaton mass terms are missing!
Analyzing Fayet-Iliopolous terms, we will see in section \ref{secfi}
that   there must exist
a stringy one-loop correction to the DUY condition.

Let us complete this section by  briefly commenting on what happens
for the case of $U(N)\times U(1)^M$ bundles.
The above discussion applies equally well to these kinds of models
 with minor changes in the concrete expressions for the various
anomalies. Defining for $m=1,\ldots,M$
\bea 
{\widehat{\ov f}^m} = \sum_{n=1}^M {\cal Q}_{mn} \, {\ov f}^n
\eea
in terms of the charge matrix (\ref{chargematrix}),
one finds in this case 
\bea
   A_{U(1)_m-SU(N)^2}&\sim &  f_m \wedge {\rm tr} F_1^2\
\left[ \int_{\cal M}  \widehat{\ov f}^m\wedge \left({\rm tr} \ov F_1^2 -
                    {1\over 2} {\rm tr} \ov R^2\right)\right] \label{GSmixedgaugeu},\\
 A_{U(1)_m-G^2_{\mu\nu}} &\sim&   f_m \wedge {\rm tr} R^2\ 
  \left[ \int_{\cal M}   \widehat{\ov f}^m \wedge  \left({12}\, {\rm tr} \ov F_1^2 - 5\,  
         {\rm tr} \ov R^2 \right)\right] \label{GSmixedgravu},
\eea
and for the cubic abelian anomalies
\bea
 A_{U(1)_m-U(1)_n-U(1)_p} &\sim& f_m\wedge  f_n\wedge  f_p\,
     \Bigl[ \int_{\cal M}   \hat c_{mnp} \,  \widehat{\ov f}^m \wedge \delta_{np} 
            \left( {\rm tr} \ov F_1^2 -
                    {1\over 2} {\rm tr} \ov R^2\right)  + \nonumber\\
         && \phantom{aaaaaaaaaaaaaaa}  
             \widehat{\ov f}^m\wedge  \widehat{\ov f}^n
            \wedge  \widehat{\ov f}^p \Bigr] \label{GSpuregaugeu}
\eea
with 
\bea
          \hat c_{mnp} = \frac{3}{8 \, \sigma_{mnp}} \, {\rm tr}_{E_8}(Q_m^2) .
\eea
Analogously to the $SU(N)\times U(1)^M$ case, these anomalies are cancelled by the generalized
Green-Schwarz mechanism, where now the matrix ${\cal Q}$
appears in appropriate places.

\subsection{Gauge kinetic function}

In this section we extract the holomorphic gauge kinetic function $f_a$ for
the $SU(N_a)$ gauge symmetries \cite{Derendinger:1985cv,Ibanez:1986xy,Choi:1985bz,Nilles:1997vk,Stieberger:1998yi}. 
Recall that up to quadratic order the four-dimensional Yang-Mills Lagrangian takes the form
\bea
       {\cal L}_{YM}={1\over 4}\, {\rm Re} (f_a)\, F\wedge\star F + {1\over 4}\,
         {\rm Im} (f_a)\, F\wedge  F. 
\eea
Dimensionally reducing the ten-dimensional tree-level term 
\bea
     S^{(10)}_{YM}={1\over 2\kappa_{10}^2} \int e^{-2\phi_{10}} \, {\alpha'\over 4}
             {\rm tr}( F\wedge\star_{10} F)
\eea
one obtains
\bea
       S^{(4)}_{YM}={1\over 2\pi} \int_{\IR_{1,3}} {\mbox{Vol}({\cal M}) \over \ell_s^6}e^{-2\phi_{10}} \, {1\over 4}
             {\rm tr}( F\wedge\star_{4} F).          
\eea
The axionic coupling is contained in (\ref{fGSc})
\bea
S^0_{GS}= {1\over 8\pi} \int_{\IR_{1,3}} b^{(0)}_0\wedge 
       {\rm tr}  (F_1\wedge F_1),
\eea       
so that the tree level gauge kinetic function is simply $f=S$ with
the complexified dilaton defined as
\bea
    S&=&{1\over 2\pi}\left[ e^{-2\phi_{10}} {{\rm Vol}({\cal M}) \over
             \ell_s^6 } + {i}\, b^{(0)}_0 \right]. 
\eea
However, there are the additional axionic couplings (\ref{fGSa})
\bea
  S_{GS}= {1\over 128\pi}
                 \, \int_{\IR_{1,3}}  \sum_{k=1}^{h_{11}}   
       \left( b^{(0)}_k \, {\rm tr}( F \wedge F) \right) 
         \,\, {\bigl({\rm tr} \ov F_1^2
         -{1\over 2} {\rm tr} \ov R^2\bigr)_k },
\eea
which are the axionic part of the  one-loop threshold corrections to
the gauge couplings.
We define the complexified K\"ahler moduli 
\bea
    T_k&=&{1\over 2\pi}\left[ -\alpha_k + {i} b^{(0)}_k \right],
\eea
where we used the expansion~(\ref{KaehlerformExpansion}).
Here $\alpha_k$ is nothing else than the normalized volume
of the two-cycle dual to the 2-form $\omega_k$. In this notation the 
volume of the Calabi-Yau is given by
\bea
      {\rm Vol}({\cal M})={1\over 6}\int_{\cal M} J\wedge J\wedge J=
     {  \ell_s^6 \over 6}\, \sum_{i,j,k} d_{ijk}\,  \alpha_i\, \alpha_j\, \alpha_k.
\eea 
Then the one-loop corrected gauge kinetic function for the non-abelian
gauge fields can be written as
\bea
     f=S+{1\over 16}\sum_k T_k\, {\bigl({\rm tr} \ov F_1^2
         -{1\over 2} {\rm tr} \ov R^2\bigr)_k}.
\eea

One can perform a similar computation for the gauge couplings of the 
abelian gauge factors. Here also the GS-term (\ref{GS3}) contributes
and the final result for the line bundles all embedded into, say, the first $E_8$ can be cast into the form
\bea
     f_{mn}&=&{\rm tr}_{E_8}(Q_m^2)\, \delta_{mn}S+{1\over 16}\sum_k {\rm tr}_{E_8}(Q_m^2)\, T_k\,
    \Bigl[
   \delta_{mn} {\bigl( {\rm tr} \ov F_1^2
         -{1\over 2} {\rm tr} \ov R^2\bigr)_k } + \nonumber \\
   && \phantom{aaaaaaaaaaaaaaaaaaaaaaaaa}
{4 \over 3}  {\rm tr}_{E_8}(Q^2_n) \,  \sum_{i,j} d_{ijk} \ov f^m_i \ov f^n_j  
\Bigr],
\eea
where also non-diagonal couplings appear. The corresponding expression for two abelian factors embedded into different $E_8$-factors reads
\bea
     f_{m_1n_2}= - {1 \over 12}   \sum_k {\rm tr}_{E_8^1}(Q_{m_1}^2)\, T_k\,  {\rm tr}_{E_8^2}(Q^2_{n_2}) \, 
 \sum_{i,j} d_{ijk} \ov f^{m_1}_i \ov f^{n_2}_j.  
\eea

Note that after including the one-loop corrections, generically
the gauge couplings for the $U(1)$ gauge factors are all
different. This is just the S-dual feature of the
non-universal gauge couplings in the intersecting D-brane
set-up \cite{Blumenhagen:2000wh}. 

\subsection{Fayet-Iliopoulos terms}
\label{secfi}

Since we are dealing  with  anomalous $U(1)$ gauge factors, there
are potential Fayet-Iliopoulos (FI) terms generated
\cite{Dine:1986zy,Dine:1987bq,Atick:1987gy,Dine:1987gj,Atick:1987qy,Dine:1987xk}.
Employing the standard supersymmetric field theory formula
\bea
\label{FIterms}
  D_m\, {\xi_m\over g_m^2}=D_m\,   {\partial {\cal K}\over \partial V_m}
\biggr\vert_{V=0},
\eea
the FI parameters $\xi_m$  can be computed from the K\"ahler potential
${\cal K}$, which in our case takes the following gauge invariant
form
\bea
{\cal K} &=&{M^2_{pl}\over 8\pi} \Biggl[   
       -\ln\Biggl(S+S^*-\sum_m Q^m_0\, V_m\Biggr)-\ln\Biggl(\sum_{i,j,k=1}^{h_{11}}
{d_{ijk}\over 6} 
\biggl(  T_i+T_i^*-\sum_m Q^m_i\, V_m\biggr) \nonumber \\
&& \phantom{aaaaaaa} \biggl(  T_j+T_j^*-\sum_m Q^m_j\, V_m\biggr)
\biggl(  T_k+T_k^*-\sum_m Q^m_k\, V_m\biggr) \Biggr) \Biggr].
\eea
Here ${M_{pl}^2\over 8\pi}=\kappa_{10}^{-2}\, e^{-2\phi_{10}}\, {\rm Vol}({\cal M})$ and $V_m$ denotes the vector superfields
\footnote{Taking only one universal K\"ahler modulus and setting $Q^m_k=0$, one recovers the familiar result~\cite{Witten:1985xb,Derendinger:1985cv}.}.
The charges $Q^m_k$
can be identified as the couplings in the mass terms (\ref{massa},\ref{massb})
using the definition
\bea
                 S_{mass}=\sum_{m=1}^M  \sum_{k=0}^{h_{11}} 
            {Q^m_k\over 2\pi\alpha'}
                \int_{\IR_{1,3}} f_m\wedge b^{(2)}_k. 
\eea
In view of (\ref{FIterms}) the FI terms are read off from the expansion of the 
K\"ahler potential  to linear order
in the gauge fields $V_m$ 
\bea
\label{fayet}
    {\xi_m\over g_m^2}&=&- {\pi \over \ell_s^6}\,
        {\rm tr}_{E_8}(Q^2_m) \Biggl[e^{-2\, \phi_{10}} {1\over 2}  \int_{\cal M} J\wedge J \wedge 
         {\ov f_m \over 2\pi}  - \nonumber \\
       && \phantom{aaaaaaaaaaaaaaaaaai} {\ell_s^4 \over 16}
       \int_{\cal M} {\ov f_m\over 2\pi} \wedge
          {\left({\rm tr} \ov F_1^2
         -{1\over 2} {\rm tr} \ov R^2\right)\over (2\pi)^2} \Biggr].
\eea
Apparently, the first term in (\ref{fayet}) arises at string tree-level, 
whereas the second term in (\ref{fayet}) is a one-loop term.
Therefore we interpret this result as evidence that there
exists  a one-loop correction to the DUY condition. 
In contrast to earlier claims  that abelian gauge fluxes freeze
some linear combinations of the K\"ahler moduli, we
now realize that actually linear combinations of the
dilaton and the K\"ahler moduli are frozen.

Let us provide independent support for this claim 
from heterotic-Type I duality\footnote{Though this argument
in the strong sense is valid for the $SO(32)$ theories, qualitatively
it should also teach us something about the $E_8\times E_8$
heterotic string.}.
It is known that a D9-brane wrapping the Calabi-Yau ${\cal M}$
with K\"ahler form $J^I$ and 
a $U(1)$ bundle with field strength
$F_m$ is supersymmetric if it
satisfies the MMMS condition \cite{Marino:1999af}
\bea
\label{MMMS}
    &&  {1\over 2}\int_{\cal M} J^I\wedge J^I \wedge {\cal F}_m -
     {1\over 3!}\int_{\cal M} {\cal F}_m \wedge {\cal F}_m \wedge {\cal F}_m
    = \\
  && \phantom{aaaaaqq}
      -  \tan\theta\left(
     {1\over 2}\int_{\cal M} {\cal F}_m \wedge {\cal F}_m \wedge J^I -
     {1\over 3!}\int_{\cal M} J^I\wedge J^I \wedge J^I \right)\nonumber
\eea
with ${\cal F}_m=2\pi \alpha'\, F_m$. 
For $\theta=0$ this looks quite similar to (\ref{fayet}), except
that the MMMS condition is at string tree level and the
${\cal F}_m^3$ term is an $\alpha'$ correction.
Applying the heterotic-Type I string duality relations 
\cite{Polchinski:1995df}
\bea
         e^{\phi_{10}^I}&=&e^{-\phi_{10}^H}, \nonumber \\
            J^I&=&J^H\,  e^{-\phi_{10}^H},
\eea  
the MMMS condition leads to
\bea
          {e^{-2\, \phi_{10}^H} \over 2}
           \int_{\cal M} J^H\wedge J^H \wedge {F}_m -
     {(2\pi \alpha')^2\over 3!}
        \int_{\cal M} {F}_m \wedge {F}_m \wedge {F}_m=0.
\eea
Qualitatively, this  has precisely the form of the abelian part 
of (\ref{fayet})
supporting our claim for the one-loop correction to
the DUY equation. A derivation of this equation via an FI term 
in an effective four-dimensional theory has been carried out
in \cite{Cremades:2002te}.

To our knowledge the inclusion
of non-abelian gauge fields and of the curvature term into the
MMMS equation (\ref{MMMS}) is not known (see for instance
\cite{Minasian:2001na,Enger:2003ue} for
some proposals). The discussion
above implies that this generalisation is likely to involve
the trace parts in (\ref{fayet}). In this respect it would
be
interesting to also compute the FI-terms for the $SO(32)$
heterotic string \cite{BHW}.

Which further corrections to the DUY condition do we have to expect? From the supergravity analysis of the D-term, it is clear that there cannot be any higher string-loop contributions, well in accord with the fact that the MMMS-condition in Type I is exact in $\alpha'$-perturbation theory. Moreover, it is known \cite{Poppitz:1998dj} that there are no one-loop Fayet-Iliopoulos terms on the Type I side. Consequently, S-duality dictates that the DUY equation is also exact in sigma-model perturbation theory. However, there might be additional non-pertrubative corrections which are beyond the scope of this paper.

Note that the corrected DUY equation has the interesting
prospect that the values of the frozen K\"ahler moduli
depend on the value of the string coupling constant and
vice versa. Of course whether such minima exist depends
on the various sign factors in (\ref{fayet}).
In particular for certain fluxes it allows one to freeze combinations of 
K\"ahler moduli and the dilaton such that both are still  
in the perturbative regime. Moreover, it is also
possible to have $U(1)$ bundles on non-degenerate Calabi-Yau manifolds
with $h_{11}=1$. In the next section we will provide an 
example which precisely shows all these features.

\section{Bundles with structure group $SU(4)\times U(1)$}

\subsection{$SU(4)$ bundles}

In the remainder of this article, we will apply the results described so far to the construction of explicit models. 
As a warm-up, in this section we consider the $E_8\times E_8$ heterotic string
compactified on a Calabi-Yau manifold ${\cal M}$ equipped with the  specific
class of bundles 
\bea
       W=V\oplus L
\eea
with structure group $G=SU(4)\times U(1)$. 
Let us first provide
the general expressions for the massless spectrum in four dimensions.

Embedding this structure group into one of the $E_8$ factors
leads to the breaking to $H=SU(5)\times U(1)_X$, where
the adjoint of $E_8$ decomposes as follows into 
$G \times H$ representations
(note that this notation is a little too sloppy since the type (i) $U(1)$ factors in
$G$ and $H$ are identical)
\bea
\label{breaking}
{\bf 248} 
 \stackrel{SU(4) \times SU(5) \times U(1)_X}{\longrightarrow}
\left\{\begin{array}{c}
({\bf 15},\1)_0 \\
(\1, \1)_0 + (\1,{\bf 10})_4 + (\1, \ov{\bf 10})_{-4} + (\1, {\bf 24})_0 \\
(\4,\1)_{-5} + (\4, \ov{\5})_3 + (\4,{\bf 10})_{-1} \\
(\ov{\4},\1)_{5} + (\ov{\4}, \5)_{-3} + (\ov{\4},\ov{\bf 10})_{1} \\
(\6,\5)_2 + (\6,\ov{\5})_{-2}
\end{array}\right\} .
\eea
As shown in Table \ref{signsa}, from (\ref{breaking}) 
one can immediately read off by which cohomology classes
the massless spectrum is determined.

\begin{table}[htb]
\renewcommand{\arraystretch}{1.5}
\begin{center}
\begin{tabular}{|c||c|}
\hline
\hline
reps. & Cohomology \\
\hline \hline
${\bf 10}_{-1}$ & $H^*({\cal M},V\otimes L^{-1})$ \\
${\bf 10}_{4}$ & $H^*({\cal M},L^{4})$ \\
\hline
$\overline \5_{3}$ & $H^*({\cal M},V\otimes L^{3})$ \\
 $\overline \5_{-2}$ & $H^*({\cal M},\bigwedge^2 V\otimes L^{-2})$ \\
\hline
 ${\bf 1}_{-5}$ & $H^*({\cal M},V\otimes L^{-5})$ \\
\hline
\end{tabular}
\caption{\small Massless spectrum of $H=SU(5)\times U(1)_X$ models.}
\label{signsa}
\end{center}
\end{table}
\noindent 
From this embedding of the structure group, we can determine the resulting
tadpole cancellation condition by computing the traces
\bea
\label{TCCSU4}
  {\rm tr}(\ov F^2)&=& \frac{1}{30}\, {\rm Tr}(\ov F^2) =  \frac{1}{30} \, \sum_{{\rm repr} \,  i } ( {\rm tr}_i {\ov F}^2_{SU(4)} + {\rm tr}_i {\ov F}^2_{U(1)})  \nonumber\\
     &=& 2\, {\rm tr}^{SU(4)}_{f}(\ov F^2_{SU(4)}) + 40\, \ov F^2_{U(1)} 
             = 4\,(2\pi)^2\, ( -c_2(V)+10\, c_1^2(L) ),\\
      {\rm tr}(\ov R^2)&=& 2\, {\rm tr}^{SU(3)}_f (\ov R^2)=-4\, (2\pi)^2\,
c_2(T), \nonumber
\eea
yielding
\bea 
\label{alf11}
      c_2(V)-10\, c^2_1(L)=c_2(T). 
\eea
Let us stress that here indeed the coefficient in front of $c_1^2(L)$
is not equal to one-half, as one might have expected. One might worry
that this is simply a consequence of the normalisation
of $U(1)_X$ in the sense that with the usual normalisation
${\rm tr}_{E_8}(Q^2)=2$ the coefficient would indeed be one.
But this is not the case, as we cannot change 
the powers of the line bundle $L$ in Table \ref{signsa}
accordingly. These powers have to be integers.
We will see in the next section that this result is consistent
with the abelian anomalies, which, due  to the general results
presented in section \ref{GSM}, must also contain 
a certain  combination of the traces in (\ref{TCCSU4}).

The net-number of chiral multiplets is given by the Euler characteristic of the
respective bundle, eq.~(\ref{Euler_character}).
Note that extra gauge bosons are counted by $H^*({\cal M},{\cal O})$, which can only 
appear in Table~\ref{signsa} if $L^4$ is the trivial bundle ${\cal O}$, 
i.e. $c_1(L)=0$.
Clearly in this case the gauge symmetry is extended to $SO(10)$, which is precisely
the commutant of $SU(4)$ in $E_8$. 
We will see that for the case that more $U(1)$ bundles are involved
the patterns of gauge symmetry enhancement are more intricate.

It is now a straightforward exercise to compute the four-dimensional
gauge anomalies.

\noindent
$\bullet$ 
The non-abelian $SU(5)^3$ anomaly is proportional to
\bea
 A_{SU(5)^3}= \chi({\cal M},V\otimes L^{-1}) + \chi({\cal M},L^{4}) -
           \chi({\cal M},V\otimes L^{3})-\chi({\cal M},{\textstyle \bigwedge^2} V\otimes L^{-2}).
\eea   
As expected just using (\ref{Euler_character}) and (\ref{wedgerel}), this anomaly
vanishes (even without invoking the tadpole cancellation condition).

\noindent
$\bullet$ 
The mixed abelian-gravitational anomaly $U(1)_X-G^2_{\mu\nu}$ however does 
not directly vanish and is given by
\bea
  A_{U(1)-G^2_{\mu\nu}}&=& -10\, \chi({\cal M},V\otimes L^{-1}) + 40\, \chi({\cal M},L^{4}) + 15\,
           \chi({\cal M},V\otimes L^{3})- \nonumber\\
    && 10\, \chi({\cal M},{\textstyle \bigwedge^2} V\otimes L^{-2})
           -5\, \chi({\cal M},V\otimes L^{-5}) \nonumber \\
    &=& 10\int_{\cal M} c_1(L)\, \left[ 12 \bigl(-c_2(V) +10\, c^2_1(L)\bigr) + 5\, c_2(T) \right].\label{AnEx1G}
\eea

\noindent
$\bullet$ Similarly the mixed abelian-non-abelian anomaly $U(1)_X-SU(5)^2$
takes the form
\bea
  A_{U(1)-SU(5)^2}&=& -3\, \chi({\cal M},V\otimes L^{-1}) + 12\, \chi({\cal M},L^{4}) + 3\,
           \chi({\cal M},V\otimes L^{3})- \nonumber \\
   && 2\, \chi({\cal M},{\textstyle \bigwedge^2} V\otimes L^{-2}) \nonumber \\
    &=& 10\int_{\cal M} c_1(L)\, \left[ 2\bigl(-\, c_2(V) +10\, c^2_1(L)\bigr) + c_2(T) \right].\label{AnEx1SU}
\eea
\noindent
$\bullet$ Finally for the $U(1)_X^3$ anomaly one obtains
\bea
  A_{U(1)^3}&=& -10\, \chi({\cal M},V\otimes L^{-1}) + 640\, \chi({\cal M},L^{4}) + 135\,
           \chi({\cal M},V\otimes L^{3})- \nonumber \\
    && 40\, \chi({\cal M},{\textstyle \bigwedge^2} V\otimes L^{-2})
           -125\, \chi({\cal M},V\otimes L^{-5}) \nonumber \\
    &=& 200\int_{\cal M} c_1(L)\, \left[ 6\bigl(- c_2(V) +10\, c^2_1(L)\bigr)+40 c^2_1(L) + 3\, c_2(T) \right].\label{AnEx1Uh3}
\eea
These results are in complete agreement with the general expressions
(\ref{GSmixedgauge} - \ref{GSpuregauge})
reviewed in section \ref{GSM} if one uses (\ref{TCCSU4}) to rewrite them in terms of traces.
Note that the integrands only vanish if $c_1(L)=0$, in which case the 
gauge group is enhanced to $SO(10)$. 

\subsection{Example: An $SU(4)$ bundle on the Quintic}

As one of its virtues, the 1-loop correction of the DUY equation  provides us with supersymmetric models inside the K\"ahler cone even for internal manifolds with $h_{11}=1$ due to the fixing of only a linear combination of the K\"ahler moduli and the dilaton. Let us illustrate this by constructing a simple though not realistic model on the mother of all Calabi-Yau compactifications, the Quintic, with Hodge numbers $(h_{21}, h_{11}) = (101, 1)$, intersection form 
\bea
I_3 = 5 \, \eta^3
\eea
and 
\bea
c_2(T) = 10\,  \eta^2.
\eea
As an example of the above construction, consider a bundle of the form
\bea
W= V_1 \oplus V_2 \oplus L,
\eea
where the  $SU(4)$ bundle $V_1$ and the line bundle $L$ are embedded into the visible 
$E_8$ factor and the $SU(4)$ bundle $V_2$ into the second $E_8$ bundle.
Concretely, we define both  vector bundles as the  cohomology of the 
monad
\bea 
\label{seq}
    0\to {\cal O}\vert_{\cal M}   \stackrel{}{\rightarrow} {\cal O}(1)^{\oplus 5} \oplus O(3) \vert_{\cal M}\to {\cal O}(8)\vert_{\cal M}\to 0
\eea
and pick $L={\cal O}(2)$. It is easy to check that this choice of data satisfies the tadpole equation
\bea
c_2(V_1) + c_2(V_2) - 10\, c_1^2(L) = c_2(T).
\eea
In particular, this implies that
\bea
\frac{1}{(2 \pi)^2}(\mbox{tr} \ov{F_1}^2 - \frac{1}{2} \mbox{tr} \ov{R}^2) = -2 c_2(T) + 4 c_2(V_2).
\eea
Upon defining the K\"ahler form $J = \ell_s^2 \, r \, \eta$, we thus arrive at the DUY equation
\bea
\label{rad}
r^2 = 10 \, e^{2 \phi_{10}},
\eea
one solution of which  indeed corresponds to the radius of the Quintic stabilized inside the 
K\"ahler cone at a value dictated by the concrete string coupling. 
It also admits  solutions where both the string coupling and the radius
are in the weakly coupled regime. Choosing for instance $g_s=0.8$ yields $r=2.52$.  
Let us stress once again 
that the naive tree level DUY equation $\int J \wedge J \wedge c_1(L) = 0$ would have 
stabilized the radius at $r=0$. 

Note also that the result (\ref{rad}) depends crucially on 
the presence of the hidden bundle $V_2$. This model is supposed to serve just as an 
illustration of the new model building possibilities and that more systematic searches may turn 
out to be fruitful. Suffice it here to merely add for completeness the not quite realistic 
chiral $SU(5)\times U(1)_X $ spectrum in Table~\ref{chirala}, where the anomalous $U(1)_X$ only survives as a global symmetry.
\begin{table}[htb]
\renewcommand{\arraystretch}{1.5}
\begin{center}
\begin{tabular}{|c||c|}
\hline
\hline
reps. & $\chi$ \\
\hline \hline
${\bf 10}_{-1}$ & $\chi({\cal M},V\otimes L^{-1}) = 290 $ \\
${\bf 10}_{4}$ & $\chi({\cal M},L^{4})= 460$ \\
\hline
$\overline \5_{3}$ & $\chi({\cal M},V\otimes L^{3}) = 170$ \\
 $\overline \5_{-2}$ & $\chi({\cal M},\bigwedge^2 V\otimes L^{-2})= 580$ \\
\hline
 ${\bf 1}_{-5}$ & $\chi({\cal M},V\otimes L^{-5})= -2150$ \\
\hline
\end{tabular}
\caption{\small Chiral spectrum of an $H=SU(5)\times U(1)_X$ model.}
\label{chirala}
\end{center}
\end{table}

\subsection{$U(4)$ bundles}

Instead of starting with an $SU(4)$ bundle, we could have also used
a $U(4)$ bundle. Such a construction has been considered in \cite{Distler:1987ee}
before. 
One starts with a bundle
\bea    \label{BundleU4}
    W=V\oplus L^{-1}, \quad {\rm with}\ c_1(V)=c_1(L), \, \, \,  \mbox{rank}(V)=4, 
\eea
which has structure group $SU(4)\times U(1)$.\footnote{Note that the two abelian factors in $V$ and $L$
are correlated via~(\ref{BundleU4}).}
This bundle $W$ can now
be embedded into an $SU(5)$ subgroup of $E_8$ so that the commutant
is again $SU(5)\times U(1)$.  
We embed the $U(1)$ bundle such that
\bea
    Q_1=(1,1,1,1,-4),
\eea
implying that the matrix ${\cal Q}$ defined in (\ref{chargematrix}) is simply
\bea
      {\cal Q}=Q_1(V)+Q_1(L)=5.
\eea
In fact, consider the breaking of the original structure group 
$SU(5) \rightarrow U(4)\times U(1)$ and the corresponding decomposition of 
$(\5,{\bf 10}) \rightarrow (\4,{\bf 10})_{-1} + (\1, {\bf 10})_4$ to read off 
the unique charge assignments of $V$ and $L$.
Consequently, the massless
spectrum is now given by the cohomology classes listed
in Table~\ref{signsb}.
\begin{table}[htb]
\renewcommand{\arraystretch}{1.5}
\begin{center}
\begin{tabular}{|c||c|}
\hline
\hline
reps. & Cohom. \\
\hline \hline
${\bf 10}_{-1}$ & $H^*({\cal M},V)$ \\
${\bf 10}_{4}$ & $H^*({\cal M},L^{-1})$ \\
\hline
$\overline \5_{3}$ & $H^*({\cal M},V\otimes L^{-1})$ \\
 $\overline \5_{-2}$ & $H^*({\cal M},\bigwedge^2 V)$ \\
\hline
 ${\bf 1}_{-5}$ & $H^*({\cal M},V\otimes L)$ \\
\hline
\end{tabular}
\caption{\small Massless spectrum of $H=SU(5)\times U(1)_X$ models.}
\label{signsb}
\end{center}
\end{table}

\noindent
The resulting tadpole cancellation condition reads
\bea
     c_2(V)-c_1^2(V)= c_2(T).   
\eea
Similarly to the former case, one can show that all non-abelian
gauge anomalies cancel and that the abelian ones,
\bea
  A_{U(1)-G^2_{\mu\nu}}
    &=& -{5\over 2}\int_{\cal M} c_1(L)\, \left[ 12\Bigl(-\, c_2(V) +\, c^2_1(L)\Bigr) + 5\,
      c_2(T) \right], \nonumber \\
  A_{U(1)-SU(5)^2}&=& 
      -{5\over 2}\, \int_{\cal M} c_1(L)\, \left[ 2\Bigl(-\, c_2(V) +\, c^2_1(L)\Bigr) + c_2(T) \right], \\
  A_{U(1)^3}&=&  -25\, \int_{\cal M} c_1(L)\, \left[ 12\Bigl(-\, c_2(V) +\, c^2_1(L)\Bigr) + 5 c^2_1(L) + 6\, c_2(T) \right],\nonumber
\eea
are cancelled
by a Green-Schwarz mechanism \footnote{Our result for the abelian anomalies
disagrees with the values given in \cite{Sharpe:1996xn}.}.
Taking into account ${\rm tr}_{E_8}(Q_1^2)=40$,  these anomalies are consistent with
the general result
(\ref{GSmixedgaugeu},\ref{GSmixedgravu},\ref{GSpuregaugeu}).

\subsection{Example:  A $U(4)$ bundle on a CICY }
\label{secex}

As a concrete example we  present a $U(4)$ bundle
on the Calabi-Yau three-fold 
\bea
         {\cal M}=\matrix{ \IP_3 \cr  \IP_1 \cr}\hskip -0.1cm\left[{\matrix{ 4\cr 2\cr}}\right] 
\eea      
with Hodge numbers $(h_{21},h_{11})=(86,2)$.
Let $\eta_1$ denote the two-form on $\IP_3$ and $\eta_2$
the two-form on $\IP_1$. The resulting Stanley-Reisner
ideal $SR=\{\eta_1^4, \eta_2^2\}$ on the ambient space
eventually determines the intersection form on the Calabi-Yau 
\bea
           I_3=2\eta_1^3 + 4\, \eta_1^2\eta_2.
\eea
Due to $\eta_2^2=0$ there are two four-forms on ${\cal M}$, 
namely $\{\eta_1^2, \eta_1\eta_2\}$.
The total Chern class of the manifold is given by
\bea
    c(T)={ (1+\eta_1)^4\, (1+\eta_2)^2 \over (1+4\eta_1+2\eta_2)},
\eea
leading in particular to
\bea 
    c_2(T)=6\eta_1^2+8\eta_1\eta_2.
\eea

To see how the  requirement of supersymmetry relates the expectation value of the dilaton and the K\"ahler moduli of the Calabi-Yau, we use the tadpole cancellation condition to arrive at $\frac{1}{(2 \pi)^2} ( \mbox{tr}({\ov F})^2 - \frac{1}{2} \mbox{tr}({\ov R})^2 ) = -2 \, c_2(T)$ for this kind of construction.   
For a general line bundle $L={\cal O}(m,n)$ and a general K\"ahler class
$J= \ell_s^2 \, ( r_1\eta_1+r_2\eta_2 ) $ with $r_{1,2}\ge 0$  the 1-loop corrected
DUY condition  reads in this case
\bea
      r_1\,  \left[(2n+m)r_1+4m\, r_2\right] + {1\over 2} ( 11m + 6n) \, e^{2 \phi_{10}} = 0,
\eea           
which leaves enough room to stabilize the ratio of the K\"ahler moduli inside the K\"ahler cone for any given value of the string coupling constant by a suitable choice of the line bundle.
The volume of the Calabi-Yau manifold is given 
by
\bea
        \mbox{Vol}({\cal M})=\frac{1}{6} \,\int_{\cal M} J^3 = \ell_s^6 \,(2\, r_1^3+12\, r_1^2 r_2).
\eea

Now we take $L={\cal O}(-2,2)$ and the $U(4)$ bundle $V$ defined
via the exact sequence
\bea
\label{extwo}
    0\! \to\! V\!   \stackrel{}{\rightarrow}\! {\cal O}(1,0)^{\oplus 2}\oplus 
            {\cal O}(0,1)^{\oplus 2} \oplus{\cal O}(1,1)^{\oplus 2}\vert_{\cal M}
     \stackrel{f}{\rightarrow} {\cal O}(4,1)\oplus {\cal O}(2,1)\vert_{\cal M}\! \to\!  0.
\eea
One can choose the map $f$ such that it does not degenerate
at any point on ${\cal M}$, which ensures that the exact sequence
(\ref{extwo}) really defines a bona-fide vector bundle. 
As with all these constructions we do not know how to proof that
the bundle is really stable. 

\begin{figure}
\begin{center}
\epsfxsize=8cm
\epsfbox{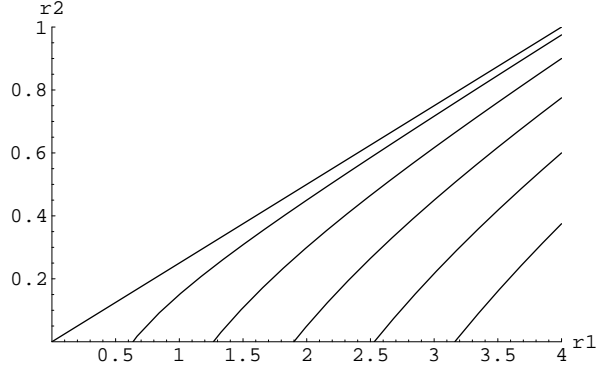}
\caption{The plot shows the K\"ahler moduli $(r_1,r_2)$ for the
values $4g_s\in\{0,0.1,0.2,0.3,0.4,0.5\}$ of the string coupling constant.}
\label{kahlaa}
\end{center}
\end{figure}

One can easily check that 
\bea
{\rm ch}_2(V)+{1\over 2}c_1^2(L^{-1})=-c_2(T)
\eea
is satisfied.
Moreover, the DUY condition imposes the constraint 
\bea
 r_1^2 - 4\, r_1 \, r_2 = {5 \over 2} \, e^{2 \phi_{10}} 
\eea
on the two radii and the dilaton.
The solutions to this constraint for certain values of the
string coupling constant are shown in Figure \ref{kahlaa}.

The next step is to compute the massless spectrum. For the chiral one
it is sufficient to compute just the various Euler characteristics,
whereas for determining also the complete non-chiral massless spectrum, 
one has to compute the various cohomology classes by tracing
through the long exact sequences in cohomology \cite{Distler:1996tj}. 
The essential input here is that the cohomology classes of line bundles
${\cal O}(m,n)$ over the ambient space $\IP_3\times \IP_1$
can be determined from Bott's formula
\bea
   h^0&=&{m+3\choose 3}(n+1)\phantom{--}\quad\quad  {\rm for}\ m,n\ge 0, \nonumber\\
   h^1&=&{m+3\choose 3}(-n-1)\phantom{-}\quad\quad {\rm for}\ m\ge 0,n\le -2, \nonumber\\
   h^2&=&0\phantom{i{m-3\choose 3}(-n-1)}\quad\quad {\rm for\ all}\ m,n, \\
   h^3&=&{-m-1\choose 3}(n+1)\phantom{-}\quad\quad {\rm for}\ m\le -4,n\ge 0,\nonumber \\           
   h^4&=&{-m-1\choose 3}(-n-1)\quad\quad {\rm for}\ m\le -4,n\le -2. \nonumber
\eea
The results for the massless spectrum are listed in Table~\ref{spectra}.

\begin{table}[htb]
\renewcommand{\arraystretch}{1.5}
\begin{center}
\begin{tabular}{|c||c|c|}
\hline
\hline
reps. & Cohom. & $\chi$  \\
\hline \hline
${\bf 10}_{-1}$ & $H^*({\cal M},V)=(0,62,0,0)$ & $-62$\\
${\bf 10}_{4}$ & $H^*({\cal M},L^{-1})=(0,10,0,0)$ & $-10$ \\
\hline
$\overline \5_{3}$ & $H^*({\cal M},V\otimes L^{-1})=(0,62,0,0)$ & $-62$  \\
 $\overline \5_{-2}$ & $H^*({\cal M},\bigwedge^2 V)=(x,10+x+y,y,0)$ & $-10$\\
\hline
 ${\bf 1}_{-5}$ & $H^*({\cal M},V\otimes L)=(0,44,62,0)$ & $18$ \\
\hline
\end{tabular}
\caption{\small Massless spectrum of an $SU(5)\times U(1)$ model.
For $H^*({\cal M},\bigwedge^2 V)$ the sequence computation is highly non-trivial 
with a vast number of  explicit maps to be investigated, and we do not perform the explicit calculation here. }
\label{spectra}
\end{center}
\end{table}
\noindent
Since $L$ is not trivial, the $U(1)_X$ symmetry becomes massive via
the Green-Schwarz mechanism but survives as a perturbative global
symmetry. Taking also this global quantum number into account
we have obtained an $SU(5)$ GUT model with $62$ generations and some 
exotic chiral states.
Ignoring the $U(1)_X$ charge, we would say that we have a model
with $72$ generations. Note that $\chi({\cal M},V)=\chi({\cal M},V\otimes L^{-1})$
is just an accident for this concrete model, but shows
that extra relations for the massless spectrum can be obtained
in this class of models. Of course such extra conditions
have to be imposed  in general to guarantee the absence of additional
exotic chiral matter.

\section{Bundles with structure group $SU(4)\times U(1)^2$}

By embedding a second $U(1)$ bundle into the observable $SU(5)$, one can break
the $SU(5)$ to the Standard Model gauge symmetry. Therefore, we now
consider  an $SU(4)\times U(1)\times U(1)$ bundle
\bea
      W=V\oplus L_1 \oplus L_2
\eea 
and a $U(4)\times U(1)\times U(1)$ bundle
\bea
      W=V\oplus L_1^{-1} \oplus L_2^{-1}
\eea 
with $c_1(W)=0$, respectively. In this latter case, the embedding
of the two $U(1)$ bundles into $SU(6)$ is given by
\bea
 Q_1=(-1,-1,-1,-1,4,0),   \quad    Q_2=(1,1,1,1,1,-5),
\eea
with ${\rm tr}_{E_8}(Q_1^2)=40$ and ${\rm tr}_{E_8}(Q_2^2)=60$.
This leads to
\bea
            {\cal Q}=\left(\matrix{ -5 & -1 \cr
                              0 & 6 \cr}\right).
\eea

\subsection{The massless spectrum and gauge enhancement}
 
The commutant in this case is $H=SU(3)\times SU(2)\times U(1)_{X}\times U(1)_{Y'}$ and
the resulting decomposition of the adjoint representation of $E_8$ reads
\bea
\label{breakingb}
{\bf 248} 
 \stackrel{SU(4) \times SU(3)\times SU(2) \times U(1)^2}{\longrightarrow}
\left\{\begin{array}{c}
({\bf 15},\1,\1)_{0,0} \\
2 \times (\1, \1, \1)_{0,0} + (\1, {\bf 8},\1)_{0,0} + (\1,\1, {\bf 3})_{0,0} \\
(\1,{\bf 3},{\bf 2})_{0,-5} + c.c. \\
(\1,{\bf 3},{\bf 2})_{4,1} + (\1,{\bf \ov 3},{\bf 1})_{4,-4} + (\1,{\bf 1},{\bf 1})_{4,6}+ c.c.\\
(\4 ,{\bf 3},{\bf 2})_{-1,1} + (\4 ,{\bf \ov 3},{\bf 1})_{-1,-4}+(\4 ,{\bf 1},{\bf 1})_{-1,6} +c.c.\\
  (\4 ,{\bf \ov 3},{\bf 1})_{3,2} + (\4 ,{\bf 1},{\bf 2})_{3,-3} + (\4 ,{\bf 1},{\bf 1})_{-5,0} +c.c\\
(\6 ,{\bf \ov 3},{\bf 1})_{-2,2} + (\6 ,{\bf 1},{\bf 2})_{-2,-3} +c.c.
\end{array}\right\}. 
\eea
Note that the $U(1)$ charges are proportional to the $U(1)_{X}$ and
$U(1)_{Y'}$ charges in the flipped $SU(5)$ GUT model. 
The (possibly anomalous)  hypercharge $U(1)_Y$ and the $U(1)_{B-L}$ charge are given by the
linear combinations
\bea
            Q_Y =-{1\over 15}\, Q_{Y'} -{2\over 5}\,  Q_{X}, \quad\quad
         Q_{B-L} = {2\over 15}\,  Q_{Y'} -{1\over 5}\, Q_{X}. 
\eea
The massless spectrum is counted by the cohomology classes in Table~\ref{signse}.
\begin{table}[htb]
\renewcommand{\arraystretch}{1.5}
\begin{center}
\begin{tabular}{|c||c|c|}
\hline
\hline
reps. & $SU(4)\times U(1)^2$ & $U(4)\times U(1)^2$ \\
\hline \hline
$(\3,\2)_{-1,1}$ & $H^*({\cal M},V \otimes L_1^{-1}\otimes L_2)$ & $H^*({\cal M},V)$ \\
$({\ov {\3}},\1)_{-1,-4}$ & $H^*({\cal M},V \otimes L_1^{-1}\otimes L_2^{-4})$ & $H^*({\cal M},V\otimes L_2^{-1})$  \\
$(\1,\1)_{-1,6}$ & $H^*({\cal M},V \otimes L_1^{-1}\otimes L_2^{6})$ & $H^*({\cal M},V\otimes L_2 )$ \\
\hline
$( \ov {\3},\1)_{3,2}$ & $H^*({\cal M},V \otimes L_1^{3}\otimes L_2^{2})$ & $H^*({\cal M},V\otimes L_1^{-1})$ \\
$(\1,\2)_{3,-3}$ & $H^*({\cal M},V\otimes L_1^{3}\otimes L_2^{-3})$  & $H^*({\cal M},V \otimes L_1^{-1}\otimes L_2^{-1})$    \\
\hline
$(\1,\1)_{-5,0}$ & $H^*({\cal M},V\otimes L_1^{-5})$   & $H^*({\cal M},V\otimes L_1)$    \\
\hline
$({\ov {\3}},\1)_{-2,2}$  & $H^*({\cal M},\bigwedge^2 V \otimes L_1^{-2}\otimes L_2^{2} )$ & $H^*({\cal M},\bigwedge^2 V)$ \\
$(\1,\2)_{-2,-3}$  & $H^*({\cal M},\bigwedge^2 V\otimes L_1^{-2}\otimes L_2^{-3} )$  & $H^*({\cal M},\bigwedge^2 V\otimes L_2^{-1} )$ \\
\hline
$(\3,\2)_{4,1}$ & $H^*({\cal M},  L_1^{4}\otimes L_2)$ & $H^*({\cal M},L_1^{-1})$  \\
$({\ov {\3}},\1)_{4,-4}$ & $H^*({\cal M}, L_1^{4}\otimes L_2^{-4})$  & $H^*({\cal M},L_1^{-1}\otimes L_2^{-1})$ \\
$(\1,\1)_{4,6}$ & $H^*({\cal M},L_1^{4}\otimes L_2^{6} )$ & $H^*({\cal M},L_1^{-1}\otimes L_2  )$  \\
$(\3,\2)_{0,-5}$ & $H^*({\cal M},L_2^{-5})$ & $H^*({\cal M},L_2^{-1})$  \\
\hline
\end{tabular}
\caption{\small Massless spectrum of $H=SU(3)\times SU(2)\times U(1)_{X}\times U(1)_{Y'}$ models.}
\label{signse}
\end{center}
\end{table}
The resulting tadpole cancellation condition reads
\bea 
      c_2(V)-10\, c^2_1(L_1)-15\, c^2_1(L_2) = c_2(T)
\eea
for the $SU(4)\times U(1)^2$ bundle and 
\bea 
      -{\rm ch}_2(V)- {1\over 2}\sum_{i=1}^2 c^2_1(L_i) = c_2(T)
\eea
for the $U(4)\times U(1)^2$ bundle.
For generic  first Chern classes $c_1(L_1)$ and $c_1(L_2)$, the two $U(1)$ gauge symmetries
are anomalous and gain a mass via the Green-Schwarz mechanism. Therefore, the
generic unbroken gauge symmetry is $SU(3)\times SU(2)$.
By computing the various anomalies, one can show that the linear combination
\bea 
       U(1)_f\simeq \kappa_1 \, U(1)_{X} + \kappa_2\, U(1)_{Y'}
\eea
is anomaly-free precisely if the first Chern classes of the two line bundles
for the $SU(4)\times U(1)^2$ case satisfy the relation
\bea 
\label{kaa}
      2\kappa_1\, c_1(L_1) + {3\kappa_2}\, c_1(L_2)   =0
\eea
and for the $U(4)\times U(1)^2$ case
\bea 
\label{kab}
   5\kappa_1\, c_1(L_1) -  (6\kappa_2-\kappa_1)\, c_1(L_2)  =0.
\eea
In the $SU(4)\times U(1)^2$ case,
for certain values of the parameters $\kappa_1,\kappa_2$ some of the line bundles 
$   L_1^{4} \otimes L_2 $, $  L_1^{4} \otimes L_2^{-4} $, $ L_1^{4} \otimes  L_2^{6} $ and
$L_2^{-5}$ appearing in Table~\ref{signse} become trivial and signal a
non-abelian  enhancement of the gauge symmetry. 
For the $U(4)\times U(1)^2$ bundles the situation is of course completely similar.
The four, respectively five for all line bundles trivial, possible
non-abelian enhancements of $SU(3)\times SU(2)$ are depicted  in Figure 1.
This shows that not only the expected $SO(10)$ and $SU(5)$ gauge groups are possible,
but also other gauge groups containing $SU(3)\times SU(2)\times U(1)^2$
as a subgroup.

Another way of understanding these gauge symmetry enhancements is 
by observing that
the linear relations (\ref{kaa},\ref{kab}) for the two line bundles imply that
the structure group is reduced to $SU(4)\times U(1)$, which of course 
enhances the  commutant. For a generic linear relation (\ref{kaa},\ref{kab})
the extra $U(1)$ appearing in the commutatant is of type (ii). 
What the commutant precisely  is, depends
on how the $U(1)$ is embedded into $SO(10)$, but
such a group theoretic  analysis is not necessary 
as one can read off the enhanced
gauge symmetries simply from Table \ref{signse}.
 
For instance, if respectively $L_1^{4}\otimes L_2^{-4}$ or $L_1^{-1}\otimes
L_2^{-1}$ become trivial, the $U(1)_{B-L}$ is anomaly-free and enhances
the $SU(3)$ gauge symmetry to $SU(4)$. Let us show a concrete example
of such a bundle. 

\begin{figure}
\begin{center}
\hbox{\epsfxsize=2.75in \epsfbox{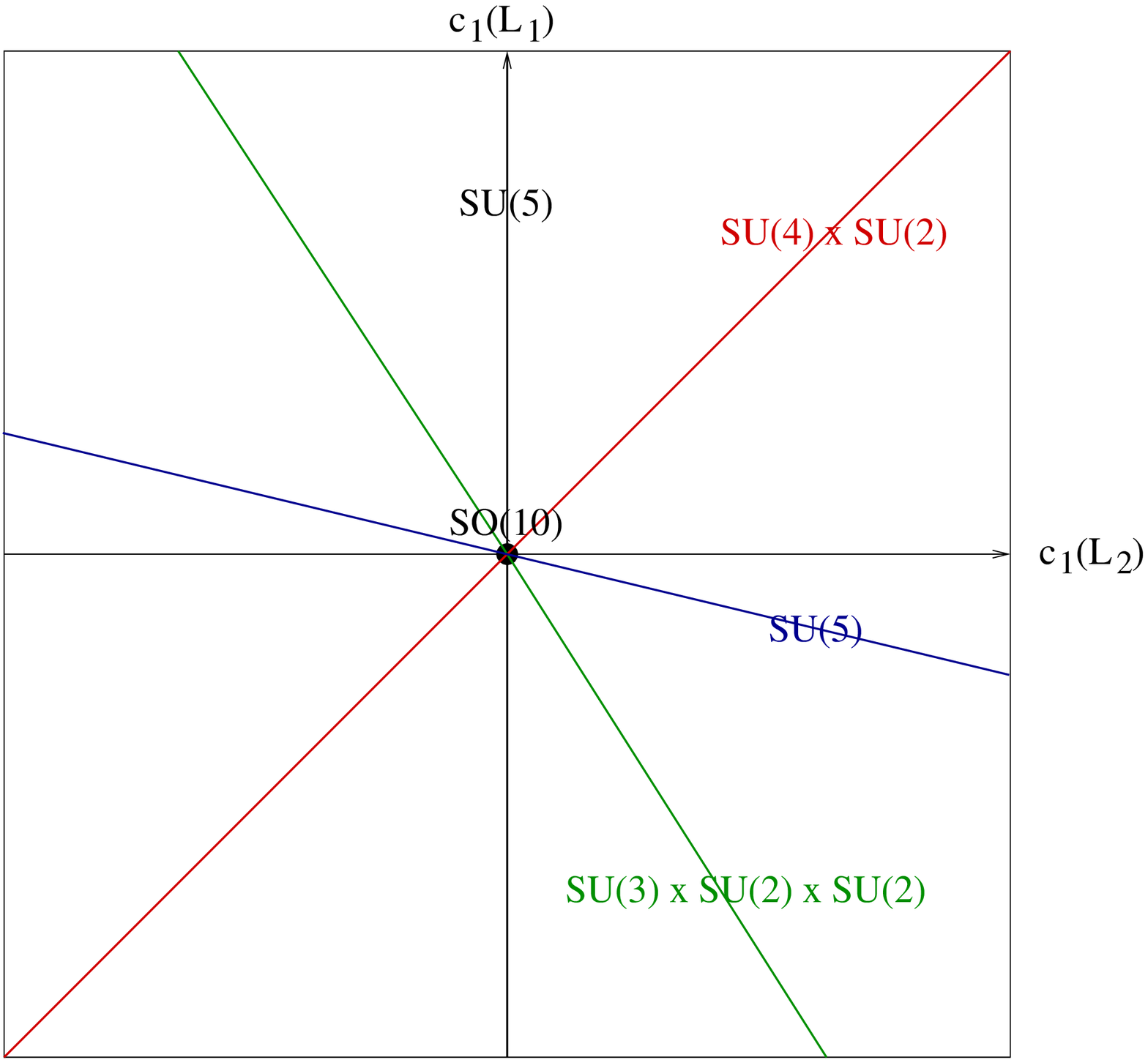} \epsfxsize=2.75in \epsfbox{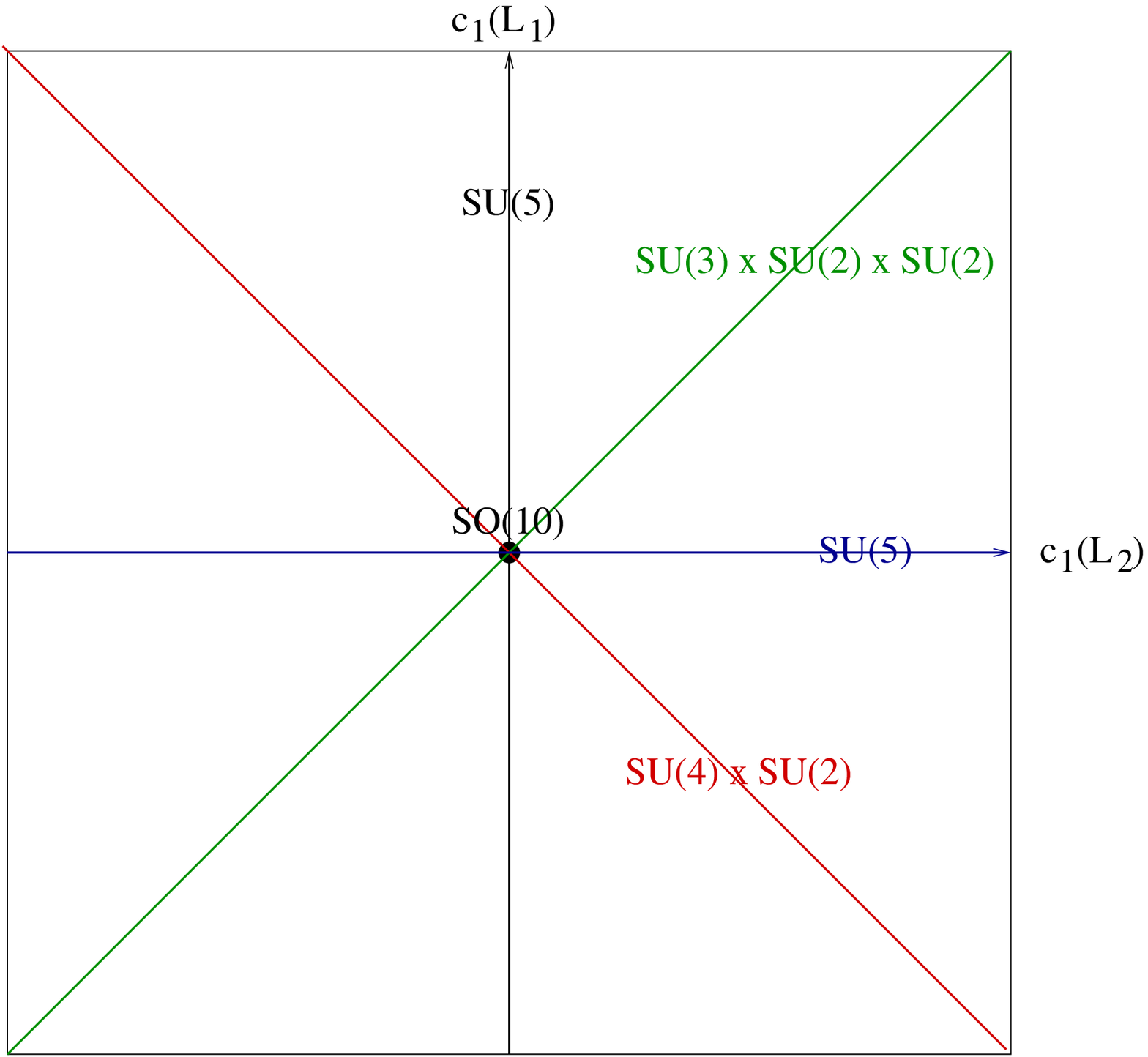}}
\caption{Gauge symmetry enhancement for bundles with structure group
$SU(4)\times U(1)^2$. On  generic lines  through the origin the gauge symmetry
is enhanced to $SU(3)\times SU(2)\times U(1)$ while for the specific
values shown one gets even non-abelian enhancement.
The left image shows the loci of non-abelian enhancement 
in the $(c_1(L_2),c_1(L_1))$-plane for $SU(4)\times U(1)^2$
and the right image for $U(4)\times U(1)^2$.
\label{enhance}}
\end{center}
\end{figure}

\subsection{Example:  A bundle with $SU(4)\times SU(2)$ gauge symmetry}

We consider  the same CICY as in section \ref{secex} and choose the two line
bundles as
\bea
      L_2=L^{-1}_1={\cal O}(-2,2).
\eea

Therefore, the K\"ahler moduli and the dilaton are again related by the constraint $r_1^2 - 4 r_1 \, r_2 = {5 \over 2} \, e^{2 \phi_{10}} $. 
From our general analysis the observable gauge symmetry is
$G=SU(4)\times SU(2)$ and one linear combination of the two $U(1)$s
is anomalous and gets a mass via the Green-Schwarz mechanism. 
In this case the $U(4)$ bundle is in fact an $SU(4)$ bundle
and one can satisfy  the tadpole cancellation condition by choosing
\bea
    0\to\ V   \stackrel{}{\rightarrow} {\cal O}(1,0)^{\oplus 5}
     \stackrel{f}{\rightarrow} {\cal O}(5,0)\vert_{\cal M}\to 0.
\eea
This bundle has the property $H^0({\cal M},V)=H^3({\cal M},V)=0$, which is a necessary
condition for the bundle to be stable.
The resulting chiral massless spectrum is listed in Table \ref{spectrb}.

\begin{table}[htb]
\renewcommand{\arraystretch}{1.5}
\begin{center}
\begin{tabular}{|c||c|}
\hline
\hline
reps. & $\chi$  \\
\hline \hline
$({\bf 4},{\bf 2})$ & $\chi({\cal M},V)=-40$ \\
$({\bf\ov 4},{\bf 1})$ & $\chi({\cal M},V\otimes L_2^{-1})=-110$ \\
$({\bf\ov 4},{\bf 1})$ & $\chi({\cal M},V\otimes L_2)=30$ \\
\hline
$({\bf 6},{\bf 2})$ & $\chi({\cal M},L_2^{-1})=-10$ \\
$({\bf 1},{\bf 2})$ & $\chi({\cal M},\bigwedge^2 V\otimes L_2^{-1})=-140$ \\
\hline
\end{tabular}
\caption{\small Massless spectrum of an $SU(4)\times SU(2)$ model.}
\label{spectrb}
\end{center}
\end{table}

Clearly, this is not a realistic model, but it shows that non-trivial
models with extra enhanced gauge symmetry can be obtained in this set-up.

\section{Bundles with structure group $SU(3)\times U(1)^3$}

\subsection{The massless spectrum and gauge enhancement}

Let us explore further the  model building possibilities several line bundles bring about and consider the embedding of a bundle of the type
\bea
W= V \oplus L_1 \oplus L_2 \oplus L_3
\eea
 with structure group $G=SU(3) \times U(1) \times U(1) \times U(1)$. We thus  
break $E_8$ down to  $H=SU(3) \times SU(2) \times U(1)_Z \times U(1)_X\times U(1)_{Y'}$ by  replacing  the internal $SU(4)$ bundle of the previous example by an $SU(3) \times U(1)_Z$ bundle. Alternatively, one can again choose 
the bundle $W$ to be of the form
\bea
W= V \oplus L_1^{-1} \oplus L_2^{-1} \oplus L_3^{-1}
\eea
and the structure group of $V$ to be $U(3)$. 
In this latter case, the embedding
of the three $U(1)$ bundles into $SU(6)$ is given by
\bea
      Q_1=(1,1,1,-3,0,0),\  Q_2=(-1,-1,-1,-1,4,0), \
       Q_3=(1,1,1,1,1,-5)
\eea
with ${\rm tr}_{E_8}(Q_1^2)=24$, ${\rm tr}_{E_8}(Q_2^2)=40$ and
${\rm tr}_{E_8}(Q_3^2)=60$.
This leads to
\bea
            {\cal Q}=\left(\matrix{ 4 & 1 & 1 \cr
                                    0 & -5 & -1 \cr
                              0 & 0  & 6 \cr}\right).
\eea
The massless spectrum for both cases is counted by the respective 
cohomology classes in Table~\ref{cohSU(3)} of Appendix B.
The resulting tadpole cancellation condition reads
\bea 
      c_2(V)-6\, c_1^2(L_1)-10\, c_1^2(L_2) - 15\, c_1^2(L_3)= c_2(T)
\eea
for the $SU(3)\times U(1)^3$ bundle and 
\bea 
      -{\rm ch}_2(V)- {1\over 2}\sum_{i=1}^3 c^2_1(L_i) 
        = c_2(T)
\eea
for the $U(3)\times U(1)^3$ bundle.

For generic  first Chern classes $c_1(L_1)$, $c_1(L_2)$  and $c_1(L_3)$ the three $U(1)$ gauge symmetries
are anomalous and gain a mass via the Green-Schwarz mechanism. Therefore, the
generic gauge symmetry is $SU(3)\times SU(2)$. However, for particular choices of the bundle data we encounter a rich pattern of gauge enhancement, as we will now discuss systematically.

The computation of the various anomalies for the $SU(3) \times U(1)^3$ case reveals that the linear combination
\bea
\label{freeuonea} 
       U(1)_f = \kappa_1 \, U(1)_{Z} + \kappa_2 \, U(1)_X + \kappa_3 \, U(1)_{Y'}
\eea
is anomaly-free precisely 
if the first Chern classes of the line bundles satisfy
\bea 6\kappa_1\, c_1(L_1) + 10\kappa_2 \, c_1(L_2) + 15\kappa_3\, c_1(L_3) = 0.
\eea
The corresponding  constraint for the $U(3)\times U(1)^3$ case reads
\bea 
\label{freeuoneb} 
      4\kappa_1 c_1(L_1)+(\kappa_1-5 \kappa_2)\, c_1(L_2) + 
              (6\kappa_3 + \kappa_1 -\kappa_2)\, c_1(L_3)  =0.
\eea
For linearly independent first Chern classes, the respective equation cannot be satisfied 
other than trivially, of course, 
and we are left with  gauge group $SU(3) \times SU(2)$. If, however, the $c_1(L_i)$ span a 
two- or one-dimensional subspace of  
their cohomology class, we can find -- modulo rescaling --  precisely one  or, respectively, 
two non-anomalous $U(1)_f$.
These $U(1)$ symmetries are of type (ii) and, as they cannot receive
a mass via GS-couplings, remain massless.

Independently of the concrete bundle data, one can check   
 that quite a few  values of $\kappa_1, \kappa_2, \kappa_3$  admit an interpretation of the corresponding abelian factor, if massless, as the MSSM hypercharge $U(1)_Y$. We list them in Table~\ref{spec1} and Table~\ref{spec2} together with the respective candidates for MSSM fermions exhibiting the required $SU(3) \times SU(2) \times U(1)_Y$ quantum numbers.

\begin{table}[htb]
\renewcommand{\arraystretch}{1.5}
\begin{center}
\begin{tabular}{|c|c||c|c|c|c|c|c|c|}
\hline
\hline
part. & class & $\left(\matrix{ \frac{1}{2}\cr  \frac{1}{10}\cr -\frac{1}{15}\cr}\right)$ & 
 $\left(\matrix{ -\frac{5}{14}\cr \frac{1}{14}\cr -\frac{13}{21}\cr}\right)$ &
$\left(\matrix{ \frac{3}{2}\cr -\frac{1}{10} \cr \frac{1}{15}\cr}\right)$ &
$\left(\matrix{ -\frac{1}{2} \cr \frac{33}{30}\cr -\frac{1}{15}\cr}\right)$ & 
$\left(\matrix{\frac{1}{2}\cr \frac{1}{2} \cr \frac{1}{3}\cr }\right)$ & 
$\left(\matrix{ \frac{1}{2}\cr -\frac{1}{10} \cr -\frac{7}{15}\cr}\right)$\\
\hline 
$Q_L$ & $D$ & $1,2,4$ & $1,3$ & $1$ & $2,3$ & $4$ & $4$  \\
$\ov{U}_R$ & $C$ &  $2,3,4$ & $4,6$ & $6,7$  & $4,7$ & $4,7$ & $4,6$  \\
$\ov{D}_R $& $C$ &  $1,5,6,7$ & $2$ & $1$ & $3$ & $1,2,5$ & $1,3,5$ \\
L & $B$ & $1,2,3,\ov{4}$ & $3 $& $\ov{4}$ &$ 2$ & $1, \ov{3}, \ov{4} $ & $1, \ov{2 }, \ov{4}$ \\
$\ov{E}_R$ & $A$ & $\ov{2},\ov{3},6 $ & $\ov{4},\ov{6}$  & $\ov{4},\ov{5}$ & $\ov{5},\ov{6} $ &  $\ov{4},5,6 $ &  $ \ov{4},5,6 $ \\
$\ov{\nu}_R$ & $A$ & $1,4,5$ & $2$ & $1$ & $3$ & $3$ & $1$  \\
\hline
\end{tabular}
\caption{\small MSSM particle candidates for choices of $(\kappa_1, \kappa_2, \kappa_3)$, part I. The labels of the representations refer to the position in the respective sections of Table~\ref{cohSU(3)} with bars denoting hermitian conjugation.}
\label{spec1}
\end{center}
\end{table}

\begin{table}[htb]
\renewcommand{\arraystretch}{1.5}
\begin{center}
\begin{tabular}{|c|c||c|c|c|c|c|c|c|}
\hline
\hline
part. & class  & $\left(\matrix{ -\frac{1}{2}\cr -\frac{1}{2}\cr \frac{1}{3}\cr}\right)$ & 
$\left(\matrix{-\frac{1}{4}\cr \frac{3}{20}\cr -\frac{4}{15}\cr}\right)$ & 
$\left(\matrix{ -1\cr \frac{1}{5}\cr -\frac{7}{15}\cr}\right)$ & 
$\left(\matrix{ -\frac{1}{12} \cr \frac{7}{60}\cr -\frac{1}{15}\cr}\right)$ & 
$\left(\matrix{ -1 \cr \frac{3}{5}\cr -\frac{1}{15}\cr}\right)$ & 
$\left(\matrix{ -\frac{1}{2}\cr \frac{7}{10}\cr -\frac{7}{15}\cr}\right)$ \\
\hline
$Q_L$ & $D$ & $4$ & $1,3$ & $1$ & $2,3$ & $2 $& $3$ \\
$\ov{U}_R$ & $C$ & $6,7$ & $5$ & $6$ & $5$ & $7$ & $4$ \\
$\ov{D}_R$ & $C$ & $2,3,5$ & $2,7$ & $4,7$ & $3,6$ & $6,4$ & $6,7$ \\
L & $B$ & $1, \ov{2}, \ov{3} $ &  $\ov{2}, \ov{4}$ & $\ov{3}, \ov{4} $&$ \ov{3}, \ov{4} $&$ \ov{2}, \ov{4}$& $\ov{2}, \ov{3}$ \\
$\ov{E}_R$ & $A$ & $4,5,\ov{6}$ & $ \ov{5}$ &$ \ov{1},2,\ov{4},\ov{5}$ & $\ov{4}$& $ 1,3,\ov{5 }$ & $\ov{2},\ov{3},\ov{6}$ \\
$\ov{\nu}_R$ & $A$ & $1$ & $ 2$ & $ 3$ & $ 3 $& $2 $& $1 $\\\hline
\end{tabular}
\caption{\small MSSM particle candidates for choices of $(\kappa_1, \kappa_2, \kappa_3)$, part II.}
\label{spec2}
\end{center}
\end{table}

\begin{figure}
\begin{center}
\epsfxsize=6.5in
\epsfbox{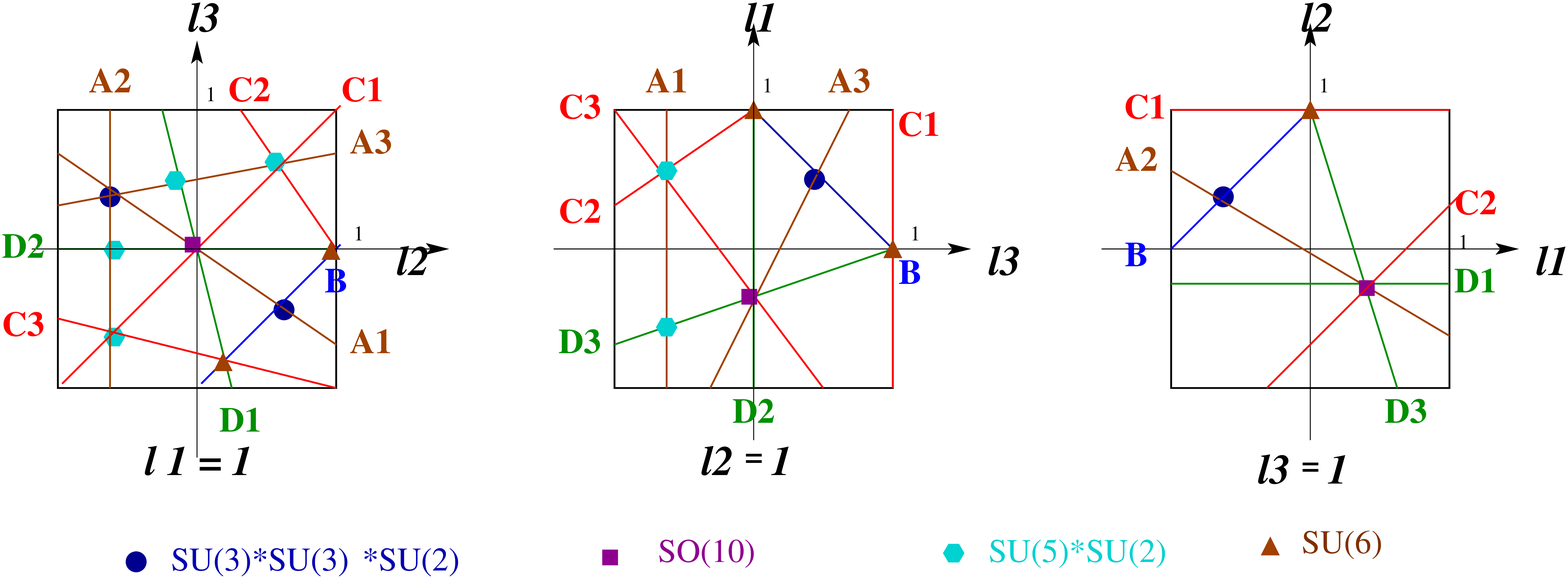}
\caption{Gauge symmetry enhancement for $SU(3)\times U(1)^3$ bundles. The picture shows the projection of the various planes defined in
 Table~\ref{enhance1} into the  planes $l_i \equiv c_1(L_i)=1$. At the point $l_i=0$ for $i=1,2,3$, the observable gauge group is $E_6$.
\label{SU(3)enhance}}
\end{center}
\end{figure}

A closer look at Table~\ref{cohSU(3)} reveals a large number of possibilities for further non-abelian gauge enhancements for those choices of $c_1(L_1), c_1(L_2), c_1(L_3)$ where additional gauge bosons in the $H^*({\cal M},\cal{O})$ representation arise. In fact, one can verify that the spectrum then organises itself into multiplets of the corresponding gauge group, as listed in Table~\ref{enhance1}. We arrive at even higher rank gauge groups if several of the states transform in the trivial bundle simultaneously. The resulting enhancement pattern is plotted schematically in Figure~\ref{SU(3)enhance} for the case that $V$ has structure group $SU(3)$. An analogous pattern can of course be derived for the $U(3)$ bundle construction.

\begin{table}[htb]
\renewcommand{\arraystretch}{1.5}
\begin{center}
\begin{tabular}{|c|c|c|c|c|}
\hline
\hline
{} & rep. & $SU(3)\times U(1)^3$ &  $U(3)\times U(1)^3$   & gauge group \\\hline
A1 & $(1,1,1)_{0,4,6}$ & $2l_2+3l_3=0$ &  $l_2-l_3=0$ & $SU(3) \times SU(2)^2$ \\
A2 & $(1,1,1)_{-3,-5,0}$ & $3l_1+5l_2=0$ & $ l_1-l_2=0$  &   $SU(3) \times SU(2)^2$ \\
A3 & $(1,1,1)_{-3,-1,6}$ & $3l_1+l_2-6l_3=0$ & $l_1-l_3=0$  & $SU(3) \times
SU(2)^2 $ \\
\hline
B1 & $(1,1, \2)_{-3,3,-3}$ & $l_1-l_2+l_3=0$ & $l_1+l_2+l_3=0$  & $SU(3) \times
SU(3)$ \\
\hline
C1 & $(1, \ov{\3},1)_{0,4,-4}$ & $l_2-l_3=0$ & $l_2+l_3=0$ & $SU(4) \times SU(2) $ \\
C2 & $(1, \ov{\3},1)_{-3,3,-2}$ & $3l_1-2l_2+3l_3=0$ & $l_1+l_2=0$  & $SU(4) \times SU(2) $ \\
C3 & $(1, \ov{\3},1)_{-3,-1,-4}$ & $3l_1+l_2+4l_3=0$ & $l_1+l_3=0$  & $SU(4) \times SU(2) $ \\\hline
D1 & $(1, \3,\2)_{0,4,1}$ & $4l_2+l_3=0$ & $l_2=0$ & $SU(5)$ \\
D2 & $(1, \3,\2)_{0,0,-5}$ & $l_3=0$ & $l_3=0$ & $SU(5) $ \\
D3 & $(1, \3,\2)_{-3,-1,1}$ & $3l_1+l_2-l_3=0$ & $l_1=0$ & $SU(5)$ \\\hline 
\end{tabular}
\caption{\small Generic enhancement of $SU(3)\times SU(2)$ by
  additional non-chiral degrees of freedom for both the $SU(3)\times U(1)^3$
   and $U(3)\times U(1)^3$ case. We use the notation $l_i=c_1(L_i)$.   
}
\label{enhance1}
\end{center}
\end{table}

\subsection{Example: A model with Standard Model gauge symmetry}

In this section we present one example our simple survey
of $U(3)\times U(1)^3$ vector bundles revealed, which shows that 
indeed the framework is rich enough to contain models
with just the Standard Model gauge symmetry (in addition
to a hidden gauge symmetry from the second $E_8$ factor). 
We consider the following Calabi-Yau
\bea
        {\cal M}=\matrix{ \IP_2 \cr  \IP_1 \cr \IP_1\cr}\hskip -0.1cm
        \left[{\matrix{ 3\cr 2\cr 2\cr}}\right] 
\eea      
with Hodge numbers $(h_{21},h_{11})=(75,3)$.
Let $\eta_1$ denote the two-form on $\IP_2$ and $\eta_2,\eta_3$
the two-forms on the two $\IP_1$ spaces. The resulting Stanley-Reisner
ideal $SR=\{\eta_1^3, \eta_2^2,\eta_3^2\}$ on the ambient space
eventually determines the intersection form on the Calabi-Yau 
\bea
           I_3=3\, \eta_1\eta_2\eta_3 + 2\, \eta_1^2\eta_2 +2\, \eta_1^2\eta_3.
\eea
Due to $\eta_2^2=\eta_3^2=0$ there are naively four four-forms on ${\cal M}$, 
namely $\{\eta_1^2, \eta_1\eta_2,\eta_1\eta_3, \eta_2\eta_3\}$, related, however, via
\bea
\label{fourrel}
      9\, \eta_1^2 -6\, \eta_1\eta_2-6\, \eta_1\eta_3 +8\, \eta_2\eta_3=0 .
\eea
Invoking the relation (\ref{fourrel})
to eliminate $\eta_1\eta_3$, the second Chern class of the manifold 
can be written as
\bea 
    c_2(T)=12\, \eta_1^2+12\,\eta_2\eta_3.
\eea
Now we choose the three line bundles on ${\cal M}$
\bea
    L_1={\cal O}(-1,1,0),\quad \quad L_2={\cal O}(-1,0,1),\quad \quad 
    L_3={\cal O}(0,-1,1),
\eea 
satisfying $c_1(L_2)=c_1(L_1)+c_1(L_3)$ and no other linear relation.
Looking back at (\ref{freeuonea}) and (\ref{freeuoneb}), one realizes
that this is consistent with the linear combination
\bea
      U(1)={1\over 2}\, U(1)_Z+ {1\over 2}\, U(1)_X + 
           {1\over 3}\, U(1)_{Y'}
\eea
being anomaly-free. Moreover, from Table \ref{spec1} we learn
that under this $U(1)$ gauge symmetry some chiral fields
have Standard Model quantum numbers. Therefore, this $U(1)$
defines a possible hypercharge $U(1)_Y$. Since there is no further
relation among the $c_1(L_i)$, the remaining two $U(1)$s receive
a mass via the Green-Schwarz mechanism and the $U(1)_Y$ as 
a type (ii) gauge factor stays massless.  

Now we define a $U(3)$ bundle with $c_1(V)=\sum_i c_1(L_i)$ by
the exact sequence
\bea
\label{exthree}
   0\to\ V \stackrel{}{\rightarrow} {\cal O}(0,1,0) \oplus 
    {\cal O}(0,0,1) \oplus {\cal O}(0,0,2) \oplus {\cal O}(1,1,1)  \vert_{\cal M}
         \stackrel{f}{\rightarrow} {\cal O}(3,2,2) \vert_{\cal M}\to 0.\nonumber\\
     \phantom{as}
\eea
Computing the second Chern classes one obtains
\bea
   -{\rm ch}_2(V)-{1\over 2}\sum_{i=1}^3 c^2_1(L_i) = 12\, \eta_1^2 +
     12\, \eta_2\eta_3 = c_2(T),  
\eea
which means that the tadpole cancellation condition is fulfilled 
right away.

It is now straightforward to compute the chiral massless spectrum
for this model. The result is shown in Appendix \ref{appc} and
exemplifies  that particles with Standard Model like quantum numbers appear,
but in addition there are some exotic chiral fields
with fractional electric charges. 
In this example the chiral exotics imply that the Standard Model
particles by themselves do not render the gauge theory anomaly-free. 
However,  one realizes that ``accidentally''  many 
chiral exotics are absent, which we think is quite encouraging
for realistic string model building. 
One could  also determine  the non-chiral matter by  computing
the various cohomology groups. We find for instance that the
complete quark-dublett spectrum reads
\bea
           H^*({\cal M},V)=(0,70,1,0)
\eea 
so that there is also one anti-generation. 
Since this model is not fully realistic anyway we skip this
elaborate computation here. 

Instead, we conclude that, 
as  for intersecting D-brane models, extra constraints 
have to be imposed to guarantee the absence of chiral exotics. 
In general of course there can appear also non-chiral exotics, 
which are visible by computing the precise cohomology classes
of all the bundles involved.

\begin{figure}
\begin{center}
\epsfxsize=8cm
\epsfbox{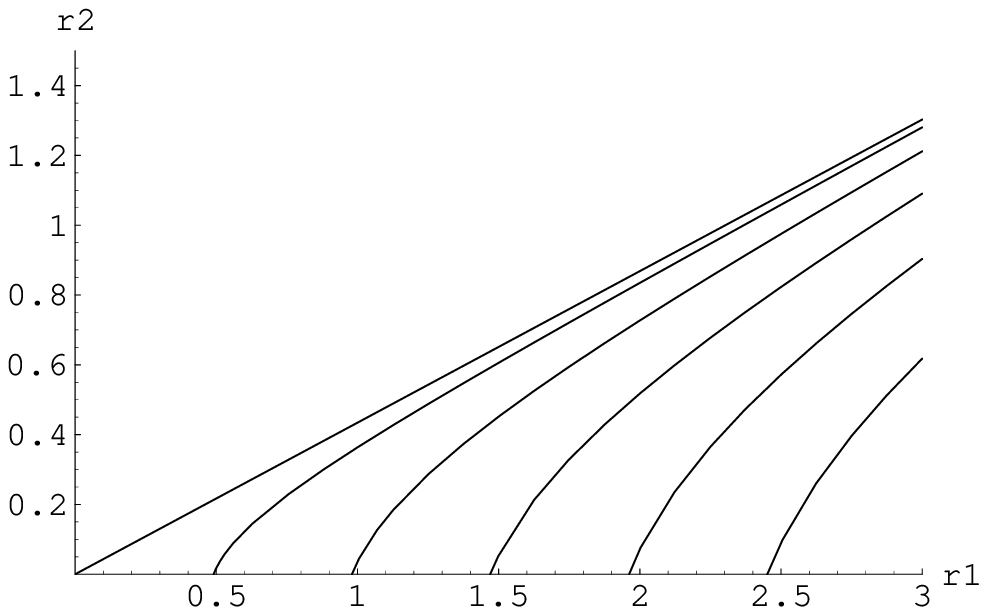}
\caption{The plot shows the K\"ahler moduli $(r_1,r_2)$ for the
values $4 g_s\in\{0,0.1,0.2,0.3,0.4,0.5\}$ of the string coupling constant.}
\label{kahlbb}
\end{center}
\end{figure}

Finally let us analyse the implications of the supersymmetry condition.
With the ansatz $J= \ell_s^2 \,   \sum_i r_i\, \eta_i$ the DUY condition reads for a general line bundle
${\cal O}(m,n,l)$
\bea
r_1 r_2 ( 3l + 2m) + r_1  r_3  (3  n + 2 m) + r_2  r_3  (3 m) + 
 r_1^2  ( n+l) = \\
  e^{2 \phi_{10}}\, (-{9 \over 2} m - 3 n - 3  l).
\eea
Our choice of bundles results in two linearly independent equations, one of which,
\bea
     r_1\, (r_2-r_3)=0,        
\eea  
dictates that we need to satisfy $r_3=r_2$ for stabilization inside the K\"ahler cone. In addition, the dilaton is related to
 the two other radii of the Calabi-Yau via
\bea
             r_1^2-3\, r_2^2 - r_1\, r_2 = {3 \over 2} \, e^{2 \phi_{10}},
\eea
which clearly admits a solution with $r_1 >0, r_2 >0$ for every fixed value of the dilaton. The solution to this equation for various values of
the string coupling constant is shown in Figure \ref{kahlbb}.

\section{Conclusions}

In this article we have investigated the model building prospects
of embedding bundles with $U(1)$ factors in their structure
group into the $E_8\times E_8$ gauge bundle of the heterotic string.
The generic features occurring are very similar to the 
Type I side. 
One encounters  anomalies for  multiple $U(1)$ gauge factors (of type (i))
which are canceled by a generalised Green-Schwarz mechanism
involving both the axio-dilaton and the axio-K\"ahler
multiplets. The induced masses for the $U(1)$ gauge bosons
can in principle be anywhere between the weak and the string scale. 
Moreover, the accompanying  FI-terms contain
the tree-level Donaldson-Uhlenbeck-Yau equation in addition to a one-loop
correction, which has just the right form to be consistent
with Type I-heterotic string duality.
The loop-corrected DUY condition freezes combinations
of the dilaton and the K\"ahler moduli. 
Again similar to Type I strings, the gauge couplings for the
$U(1)$ gauge factors of type (i) are non-universal if one includes
the one-loop threshold corrections. 

Concerning concrete model building, we have demonstrated that it is possible 
to break the ten-dimensional gauge symmetry directly
to the Standard Model gauge group. 
For special combinations of line bundles the
models experience non-abelian gauge symmetry enhancements,
which for many $U(1)$ factors show a rich pattern  of
possible gauge groups. 
As on the Type I side, new (chiral) exotic fields with 
non-Standard Model quantum numbers do arise, whose
absence imposes extra constraints on the respective bundle
cohomology classes.

Using exact sequences of sums of line bundles, 
we have also constructed a number of concrete examples showing
that indeed bundles of the described type can be found. 
The models presented here are not yet realistic, but we do
not see any conceptual obstacle to obtaining more realistic
heterotic string models of this type. Of course, for concreteness
we studied just a very few possible embeddings of 
$U(1)$ structure groups into $E_8\times E_8$. As with
intersecting D-branes, many more embeddings are possible and
some of them might turn out to be of  phenomenological
interest. 

From the phenomenological point of view, there are a couple
of open questions. We have not addressed  Yukawa couplings
in this paper and it would be interesting to see
whether they can give rise to a hierarchy of fermion masses. 
Moreover, one could try to freeze some of the
complex structure and bundle moduli by turning
on some background H-flux \cite{Strominger:1986uh,Becker:2002sx,Cardoso:2002hd}. 
As on the Type IIB side, a
supersymmetry breaking flux is expected to
induce soft terms in the observable gauge sector.
The analogous story for the $SO(32)$ heterotic string will be
slightly different, as there exists an independent fourth order
Casimir. We plan to address some of these issues in \cite{BHW}.

On the more formal side, the question arises  whether  
these models also admit  a linear sigma model \cite{Witten:1993yc} respectively
Landau-Ginzburg \cite{Distler:1993mk} or  conformal field theory
description \cite{Blumenhagen:1995tt}. This would be 
interesting, as it is by now well established that such
models are not destabilized by world-sheet instantons
\cite{Silverstein:1995re,Basu:2003bq,Beasley:2003fx}.
Moreover, one could look for 
perturbative dualities among such models like $(0,2)$ mirror
symmetry 
\cite{Blumenhagen:1996vu,Blumenhagen:1996tv,Sharpe:1998wa,Adams:2003zy} 
or the dualities described in \cite{Distler:1995bc,Blumenhagen:1997vt}.

Last but not least, these heterotic string vacua constitute an additional
branch in the string theory landscape and one might try to invoke 
statistical methods to get estimates on the
distributions of various physical quantities in the gauge theory sector
\cite{Douglas:2003um,Kumar:2004pv,Blumenhagen:2004xx,Arkani-Hamed:2005yv}.

 \vskip 1cm
 {\noindent  {\Large \bf Acknowledgments}}
\vskip 0.5cm  We thank Volker Braun, Peter Mayr and Stephan Stieberger for very helpful
discussions. R.B. would like to thank the Fields Institute in Toronto 
and the Perimeter Institute in  Waterloo
for hospitality during part of this work.

\vskip 2cm

\appendix
\section{Computing cohomology classes}

In general a short exact sequence of sheaves 
\bea
\label{short}
0 \to A \buildrel \alpha \over \to B \buildrel \beta \over 
        \to C \to 0
\eea
implies a long exact sequence in cohomology
\bea
\label{lesq}
   0 \to H^0(M,A) \buildrel \alpha \over \to
                 H^0(M,B) \buildrel \beta \over \to
                 H^0(M,C) \buildrel \phi \over \to
                 H^1(M,A) \buildrel \alpha \over \to
                 H^1(M,B) \to\ldots.
\eea
The maps $\alpha$ and $\beta$ in (\ref{lesq}) are induced from the sheaf
homomorphisms in (\ref{short}). For the definition of $\phi$ we refer to
the mathematical literature, but it is emphasised 
that the definition of $\phi$ relies on the shortness of
the sequence (\ref{short}).

In order to use the long exact cohomological sequences (\ref{lesq}) in our case, 
one  has to know the cohomology classes of line bundles restricted
to the complete intersection locus defining the Calabi Yau 3-fold. To this end one uses the Koszul
sequence for a complete intersection of $K$ hypersurfaces
$\xi=(f_1,\ldots,f_K)$ with $f_i$ a section of the line bundle
${\cal E}_{f_i}$ over the ambient space
\bea
\label{koszul}
 0\to \wedge^K {\cal E}^* \buildrel \xi\cdot\over \to \ldots
             \buildrel \xi\cdot\over \to \wedge^2 {\cal E}^*
             \buildrel \xi\cdot\over \to {\cal E}^*
             \buildrel \xi\cdot\over \to {\cal O}
             \buildrel \rho\cdot\over \to {\cal O}\vert_{\cal M} \to 0 .
\eea
Here ${\cal E}=\bigoplus {\cal E}_{f_i}$, ${\cal O}$ denotes 
the structure sheaf of the ambient space and $\rho$ is the
restriction map. 

For computing the cohomology of anti-symmetric products of bundles, a
useful fact is that a short exact sequence (\ref{short}) 
implies the following set of exact sequences\footnote{We thank
Volker Braun for communicating this mathematical fact to us.}
\bea
              \vcenter{
                \hbox{$\phantom{0\to \bigwedge^2 A \to i} 0 
                       \phantom{aaai\to } 0
                      \phantom{\to 0}$}
                 \hbox{$\phantom{0\to \bigwedge^2 A \to ,} \downarrow 
                      \phantom{aai\to } \downarrow
                      \phantom{\to 0}$}
                \hbox{$0\to \bigwedge^2 A \to Q_1\to A\otimes C\to 0$}
                \hbox{$\phantom{0\to \bigwedge^2 A \to ,} \downarrow 
                      \phantom{aai\to } \downarrow
                      \phantom{\to 0}$}
                \hbox{$0\to \bigwedge^2 A \to \bigwedge^2 B \to Q_2\to 0$}
                \hbox{$\phantom{0\to \bigwedge^2 A \to ,} \downarrow 
                      \phantom{aai\to } \downarrow
                      \phantom{\to 0}$}
                \hbox{$\phantom{0\to \bigwedge^2 A \to } \bigwedge^2 C 
                      \phantom{\to } \bigwedge^2 C
                      \phantom{\to 0}$}
                \hbox{$\phantom{0\to \bigwedge^2 A \to ,} \downarrow 
                      \phantom{aai\to } \downarrow
                      \phantom{\to 0}$}
                 \hbox{$\phantom{0\to \bigwedge^2 A \to i} 0 
                       \phantom{aaai\to } 0
                      \phantom{\to 0}$}}
\eea
which for $C$ being a line bundle reduces to 
the following short exact sequence 
\bea
\label{shortwedge}
0 \to {\textstyle\bigwedge^2} A \to {\textstyle \bigwedge^2} B 
        \to A\otimes C \to 0.
\eea

\section{The general massless spectrum for $SU(3)\times U(1)^3$ bundles}

The massless spectrum is counted by the cohomology classes in Table~\ref{cohSU(3)}.
\begin{table}[htb]
\renewcommand{\arraystretch}{1.5}
\begin{center}
\begin{tabular}{|c|c||c|c|}
\hline
\hline
class & reps. & $SU(3)\times U(1)^3$ & $U(3)\times U(1)^3$ \\
\hline \hline
$D1$ & $(\1;\3,\2)_{0,4,1}$ & $H^*({\cal M}, L^{4}_2\otimes L^{1}_{3})$ & $H^*({\cal M},L_2^{-1})$ \\
$D2$ & $(\1;\3,\2)_{0,0,-5}$ & $H^*({\cal M}, L^{-5}_{3})$ & $H^*({\cal M},L_3^{-1})$ \\
$D3$ &$(\1;\3,\2)_{-3,-1,1}$ & $H^*({\cal M}, L^{-3}_1\otimes L^{-1}_{2}\otimes L^{1}_3)$ & $H^*({\cal M},L_1^{-1})$ \\
$D4$ & $(\3;\3,\2)_{1,-1,1}$ & $H^*({\cal M},V\otimes L^{1}_1\otimes L^{-1}_{2} \otimes L^{1}_3)$ & $H^*({\cal M},V)$ \\
\hline
$B1$ & $(\1;\1,\2)_{-3,3,-3}$ & $H^*({\cal M}, L^{-3}_1\otimes L^{3}_{2} \otimes L^{-3}_3)$ & $H^*({\cal M},L_1^{-1}\otimes L_2^{-1}\otimes L_3^{-1})$ \\
$B2$ & $(\3;\1,\2)_{-2,-2,-3}$ & $H^*({\cal M},V\otimes L^{-2}_1\otimes L^{-2}_{2} \otimes L^{-3}_3)$ & $H^*({\cal M},V\otimes L_1^{-1}\otimes L_3^{-1})$ \\
$B3$ & $(\3;\1,\2)_{-2,2,3}$ & $H^*({\cal M},V\otimes L^{-2}_1\otimes L^{2}_{2} \otimes L^{3}_3) $  & $H^*({\cal M},V\otimes L_1^{-1}\otimes L_2^{-1})$ \\
$B4$ & $(\3;\1,\2)_{1,3,-3}$ & $H^*({\cal M},V\otimes L^{1}_1\otimes L^{3}_{2} \otimes L^{-3}_3) $& $H^*({\cal M},V\otimes L_2^{-1}\otimes L_3^{-1})$ \\
\hline
$C1$ & $(\1;\ov {\3},\1)_{0,4,-4}$ & $H^*({\cal M}, L^{4}_{2} \otimes L^{-4}_3)$ & $H^*({\cal M},L_2^{-1}\otimes L_3^{-1})$ \\
$C2$ & $(\1;\ov {\3},\1)_{-3,3,2}$ & $H^*({\cal M}, L^{-3}_1\otimes L^{3}_{2}  \otimes L^{2}_3  )$ & $H^*({\cal M},L_1^{-1}\otimes L_2^{-1})$ \\
$C3$ & $(\1;\ov {\3},\1)_{-3,-1,-4}$ & $H^*({\cal M}, L^{-3}_1\otimes L^{-1}_{2}\otimes L^{-4}_3)$ & $H^*({\cal M},L_1^{-1}\otimes L_3^{-1})$ \\
$C4$ & $(\3;\ov {\3},\1)_{-2,-2,2}$ & $H^*({\cal M},V\otimes L^{-2}_1\otimes L^{-2}_{2}\otimes L^{2}_3)$ & $H^*({\cal M},V\otimes L_1^{-1})$ \\
$C5$ & $(\ov {\3};\ov {\3},\1)_{2,-2,2}$ & $H^*({\cal M},\bigwedge^2 V\otimes L^{2}_1\otimes L^{-2}_{2}\otimes L^{2}_3)$ & $H^*({\cal M},\bigwedge^2 V)$ \\
$C6$ & $(\3;\ov {\3},\1)_{1,3,2}$ & $H^*({\cal M},V\otimes L^{1}_1\otimes L^{3}_{2}\otimes L^{2}_3)$ & $H^*({\cal M},V\otimes L_2^{-1})$ \\
$C7$ & $(\3;\ov {\3},\1)_{1,-1,-4}$ & $H^*({\cal M},V\otimes L^{1}_1\otimes L^{-1}_{2} \otimes L^{-4}_3)$ & $H^*({\cal M},V\otimes L_3^{-1})$ \\
\hline
$A1$ & $(\1;\1 ,\1)_{0,4,6}$ & $H^*({\cal M}, L^{4}_{2}\otimes L^{6}_3)$ & $H^*({\cal M},L_2^{-1}\otimes L_3)$ \\
$A2$ & $(\1;\1 ,\1)_{-3,-5,0}$ & $H^*({\cal M}, L^{-3}_1\otimes L^{-5}_{2})$ & $H^*({\cal M},L_1^{-1}\otimes L_2)$ \\
$A3$ & $(\1;\1 ,\1)_{-3,-1,6}$ & $H^*({\cal M}, L^{-3}_1\otimes L^{-1}_{2}\otimes L^{6}_3)$ & $H^*({\cal M},L_1^{-1}\otimes L_3)$ \\
$A4$ & $(\3;\1,\1)_{1,-5,0}$ & $H^*({\cal M},V\otimes L^{1}_1\otimes L^{-5}_{2})$ & $H^*({\cal M},V\otimes L_2)$ \\
$A5$ & $(\3;\1,\1)_{1,-1,6}$ & $H^*({\cal M},V\otimes L^{1}_1\otimes L^{-1}_{2}\otimes L^{6}_3)$ & $H^*({\cal M},V\otimes L_3)$ \\
$A6$ & $(\3;\1,\1)_{4,0,0}$ & $H^*({\cal M},V\otimes L^{4}_1)$ & $H^*({\cal M},V\otimes L_1)$ \\
\hline \hline
\end{tabular}
\caption{\small Massless spectrum of $H=SU(3)\times SU(2)\times U(1)^3$ models.}
\label{cohSU(3)}
\end{center}
\end{table}

\section{The massless spectrum for the $SU(3)\times SU(2)\times U(1)_Y$ example}
\label{appc}

In Table~\ref{cohSM} we list the chiral massless spectrum of the 
model discussed in section 5.2 with Standard Model gauge symmetry.

\begin{table}[htb]
\renewcommand{\arraystretch}{1.5}
\begin{center}
\begin{tabular}{|c|c|c|c|c|c|}
\hline
\hline
part. & reps. & class & $\chi$ & $\chi_{tot}$ & $Q_Y$  \\
\hline \hline
$Q_L$ & $(\3,\2)_{1,-1,1}$ & $D4$ & $-69$ & $-69$ &  ${1\over 3}$\\
\hline
$U_R$ & $(\ov {\3},\1)_{-2,-2,2}$ & $C4$ & $-75$ & & $ -{4\over 3}$ \\
$U_R$ & $(\ov {\3},\1)_{1,-1,4}$ & $C7$ & $-63$  & $-138$ & $ -{4\over 3}$ \\
\hline
$D_R$ & $(\ov {\3},\1)_{0,4,-4}$ & $C1$ & $-6$ & & $ {2\over 3}$ \\
$D_R$ & $(\ov {\3},\1)_{-3,3,2}$ & $C2$ & $0$  & & $ {2\over 3}$ \\
$D_R$ & $(\ov {\3},\1)_{2,-2,2}$ & $C5$ & $75$  & $69$ & $ {2\over 3}$ \\
\hline
$L_L$ & $({\1},\2)_{-3,3,-3}$ & $B1$ & $-6$  & & $ {-1}$ \\
$L_L$ & $({\1},\2)_{2,-2,-3}$ & $\ov{B3}$ & $75$ &  & $ {-1}$ \\
$L_L$ & $({\1},\2)_{-1,-3,3}$ & $\ov{B4}$ & $69$  & $138$ & $ {-1}$ \\
\hline
$E_R$ &  $({\1},\1)_{-1,5,0}$ & $\ov{A4}$ & $57$  & & $ {2}$ \\
$E_R$ & $({\1},\1)_{1,-1,6}$ & $A5$ & $-63$  & & $ {2}$ \\
$E_R$ & $({\1},\1)_{4,0,0}$ & $A6$ & $-63$  & $-69$ & $ {2}$ \\
\hline
$N_R$ & $({\1},\1)_{-3,-1,6}$ & $A3$ & $-6$  & $-6$ &  $ {0}$ \\
\hline
\hline
$Q^{ex}_L$ & $(\3,\2)_{0,4,1}$ & $D1$ & $0$ & &${7\over 3}$\\
$Q^{ex}_L$ & $(\3,\2)_{0,0,-5}$ & $D2$ & $0$ & &$-{5\over 3}$\\
$Q^{ex}_L$ & $(\3,\2)_{-3,-1,1}$ & $D3$ & $0$ & $0$ & $-{5\over 3}$\\
\hline
$Q^{ex}_R$ & $(\ov {\3},\1)_{-3,-1,-4}$ & $C3$ & $0$ && $ -{10\over 3}$ \\
$Q^{ex}_R$ & $(\ov {\3},\1)_{1,3,2}$ & $C6$ & $-69$ & $-69$ &  $ {8\over 3}$ \\
\hline
$L^{ex}_L$ & $({\1},\2)_{-2,-2,-3}$ & $B2$ & $-69$ & $-69$  & $ {-3}$ \\
\hline
$L^{ex}_R$ & $({\1},\1)_{0,4,6}$ & $A1$ & $0$  & & $ {4}$ \\
$L^{ex}_R$ & $({\1},\1)_{3,5,0}$ & $\ov{A2}$ & $0$  & $0$ & $ {4}$ \\
\hline \hline
\end{tabular}
\caption{\small Massless spectrum of $H=SU(3)\times SU(2)\times U(1)_Y$ model.}
\label{cohSM}
\end{center}
\end{table}

\clearpage
\nocite{*}
\bibliography{rev}
\bibliographystyle{utphys}

\end{document}